\documentclass[12pt,pre,eqsecnum,aps,epsfig,floats]{revtex4}

\usepackage{graphics}
\usepackage{epsfig}
\usepackage{amsfonts}
\usepackage{amsmath}
\usepackage{theorem}

\newcommand{\one}{\mbox{\tt 1}\hspace{-0.057 in}\mbox{\tt l}}

\newcommand{\Ket}[1]{|#1\rangle}

\newcommand{\x}{\mbox{\boldmath $x$}}
\newcommand{\xs}{\mbox{\boldmath $\scriptstyle x$}}
\newcommand{\y}{\mbox{\boldmath $y$}}
\newcommand{\ys}{\mbox{\boldmath $\scriptstyle y$}}
\newcommand{\z}{\mbox{\boldmath $z$}}
\newcommand{\zs}{\mbox{\boldmath $\scriptstyle z$}}
\newcommand{\as}{\mbox{\boldmath $\scriptstyle a$}}
\newcommand{\at}{\mbox{\boldmath $a$}}
\newcommand{\bs}{\mbox{\boldmath $\scriptstyle b$}}
\newcommand{\bt}{\mbox{\boldmath $b$}}
\newcommand{\xibold}{\mbox{\boldmath $\xi$}}
\newcommand{\xibolds}{\mbox{\boldmath $\scriptstyle \xi$}}
\newcommand{\gammas}{\mbox{$\scriptstyle \gamma$}}
\newcommand{\boldarrow}{\mbox{\boldmath $\swarrow$}}

\newcommand{\Tr}{\mbox{\rm\small Tr\ }}
\newcommand{\balpha}{\mbox{\boldmath $\alpha$}}
\newcommand{\balphas}{\mbox{\boldmath $\scriptstyle \alpha$}}

\newcommand{\bbetas}{\mbox{\boldmath $\scriptstyle \beta$}}

\newcommand{\bxi}{\mbox{\boldmath $\xi$}}
\newcommand{\bxis}{\mbox{\boldmath $\scriptstyle\xi$}}
\newcommand{\boldeta}{\mbox{\boldmath $\eta$}}
\newcommand{\boldetas}{\mbox{\boldmath $\scriptstyle\eta$}}
\newcommand{\bzeta}{\mbox{\boldmath $\zeta$}}
\newcommand{\bzetas}{\mbox{\boldmath $\scriptstyle\zeta$}}

\theoremstyle{break}

\begin{document}

\title{Classical predictability and coarse-grained      
evolution of the quantum baker's map}
\author{Artur Scherer}
\author{Andrei N. Soklakov}
\author{R\"udiger Schack}
\affiliation{Department of Mathematics, 
Royal Holloway, University of London,
Egham, Surrey TW20 0EX, UK}

\date{\today}

\begin{abstract}
  We investigate how classical predictability of the coarse-grained evolution
  of the quantum baker's map depends on the character of the coarse-graining.
  Our analysis extends earlier work by Brun and Hartle [Phys.\ Rev.\ D {\bf
    60}, 123503 (1999)] to the case of a chaotic map. To quantify 
  predictability, we compare the rate of entropy increase for a family of
  coarse-grainings in the decoherent histories formalism. We find that the
  rate of entropy increase is dominated by the number of scales characterising
  the coarse-graining.
\end{abstract}

\maketitle

\section{Introduction}

The concept of coarse-graining plays an important role in the emergence of
classical evolution from the fundamental quantum-mechanical equations of
motion~\cite{Gell-MannHartle1993,BrunHartle1999-PRD}.  The form of the
effective classical equations of motion is as much influenced by the character
of the coarse-graining as by the fundamental quantum-mechanical equations of
motion themselves.  A systematic way to study coarse-grained quantum evolution
is provided by the decoherent histories
formalism~\cite{Griffiths1984,Omnes1988,Gell-MannHartle1990,DowkerHalliwell1992,Gell-MannHartle1993}.
Within this approach to quantum theory a quantum mechanical system is said to
exhibit classical behaviour when histories with correlations in time that are
implied by classical deterministic laws have high
probability~\cite{Gell-MannHartle1993,BrunHartle1999-PRD}.

Coarse-grained descriptions are also used in classical physics to reduce the
number of variables when the number of degrees of freedom is large. This leads
to effective equations of motion for the coarse-grained variables.  The
character of the coarse-graining is important here. Although a given physical
system may be described by many alternative sets of coarse-grained variables,
some coarse-grained descriptions are more useful for prediction than others.
For a practical set of coarse-grained variables, the observables of interest
should be simple and slowly varying functions.

In quantum theory, the nonuniqueness of the coarse-graining procedure motivates
this question: what distinguishes coarse-grainings leading to predictable,
deterministic effective classical evolution from other coarse-grainings?  In
general, arbitrarily many sets of alternative coarse-grained histories
decohere and so can be assigned probabilities.  Moreover, two such decoherent
sets of histories are in general mutually incompatible.  Which of these many
possible coarse-grainings lead to predictable evolution of the coarse-grained
variables, i.e., useful regularities in time governed by effective,
phenomenological equations of motion?  

These questions have been addressed by Brun and Hartle in
Ref.~\cite{BrunHartle1999-PRD}, where they investigate the origin of classical
predictability by considering the simplest linear system with a continuum
description---the linear one-dimensional harmonic chain regarded as a closed
quantum mechanical system.  In their analysis a chain of $\mathcal{N}$ atoms
is divided up into groups of $N$ atoms each. Each such group is then itself
further subdivided into $N/d$ equally spaced clumps of $d$ atoms each, with a
distance between clumps of $(\mathcal{N}/N)\cdot d$.  A family of
coarse-grained descriptions is introduced by restricting attention to the
average positions of the atoms in a group, which are regarded as the relevant
variables defining the system under consideration, and ignoring the internal
coordinates within each group, which are regarded as the ``environment''. In
the case $d=N$ the $N$ atoms of each group are all neighbours. The
corresponding coarse-grained description is therefore entirely local. As $d$
decreases from $N$ to $1$ the coarse-grained description becomes more and more
nonlocal. In the case $d=1$ the $N$ atoms of each group are dispersed over
the whole chain. Brun and Hartle analyse how decoherence, noise and
computational complexity of the coarse-grained evolution depends on the
nonlocality parameter $d$ and thus show that local coarse-grainings are
characterised by a higher degree of classical predictability.

The dynamical system studied by Brun and Hartle is linear. In this paper we
analyse classical predictability for a family of coarse grainings for a {\em
  nonlinear\/} chaotic map, the quantum baker's
map~\cite{Balazs1989,Saraceno1990}.  To quantify predictability, we compute
the entropy increase for the evolution: the greater the rate of entropy
increase, the less predictable is the evolution. We consider a family of 
hierarchical multi-scale coarse grainings and show that predictability
decreases as the number of scales characterising the coarse-graining increases.

The paper is organised as follows.  We start with a short introduction
to the quantum baker's map (Sec.~\ref{sec:baker}) and the decoherent histories
formalism (Sec.~\ref{sec:DecHist}). We then introduce the family of coarse grainings
(Sec.~\ref{sec:DifferentCoarse-grainedDescriptions}), describe how the rate
of entropy increase depends on the coarse-graining (Sec.~\ref{sec:Results}),
and finally present detailed derivations of our results
(Sec.~\ref{sec:Derivations}).
 
\section{Background}
\subsection{Quantum baker's map} \label{sec:baker}

The quantum baker's map \cite{Balazs1989,Saraceno1990} is a prototypical
quantum map invented for the theoretical investigation of quantum chaos. 
It was introduced as a quantised version of the classical
baker's transformation \cite{Arnold1968}. There is, however, no unique
quantisation procedure \cite{Berry1979}. The original definition of the
map~\cite{Balazs1989,Saraceno1990} is based on Weyl's
quantisation~\cite{Weyl1950} of the unit square. In
\cite{Schack2000a} a class of quantum baker's maps has been defined by
exploiting formal similarities between the symbolic dynamics
\cite{Alekseev1981} for the classical baker's map on the one hand and the
dynamics of strings of quantum bits (qubits) on the other hand. These maps
admit a symbolic description in terms of shifts on strings of qubits similar
to classical symbolic dynamics~\cite{Alekseev1981}. Their symbolic description
has been further developed in~\cite{Soklakov2000a}.

Let us give a short introduction following~\cite{Schack2000a}.  
Quantum baker's maps are defined on the 
$D$-dimensional Hilbert space of the quantised unit square \cite{Weyl1950}. 
For consistency of units, we let the quantum scale on ``phase space''  be 
$2\pi\hbar=1/D$. Following Ref.~\cite{Saraceno1990}, we choose half-integer 
eigenvalues $q_j=(j+{1\over2})/D$, $j=0,\ldots,D-1$, and $p_k=(k+{1\over2})/D$,
$k=0,\ldots,D-1$, of the discrete ``position'' and ``momentum''
operators $\hat q$ and $\hat p$, respectively, corresponding to
antiperiodic boundary conditions.  We further assume that $D=2^N$,
which is the dimension of the Hilbert space of $N$ qubits.

The $D=2^N$ dimensional Hilbert space modelling the unit square can be 
identified with the product space of $N$ qubits via
\begin{equation}
\Ket{q_j} =
\Ket{\xi_1}\otimes\Ket{\xi_2}\otimes\cdots\otimes\Ket{\xi_N}  \;,
\label{eq:tensor1}
\end{equation}
where $j=\sum_{l=1}^N \xi_l2^{N-l}$, $\xi_l\in\{0,1\}$,
and where each qubit has basis states $|0\rangle$ and $|1\rangle$.
We can write $q_j$ as a binary fraction, $q_j=0.\xi_1\xi_2\ldots\xi_N1$. 
Let us define the notation 
\begin{equation}
\Ket{.\xi_1\xi_2\ldots \xi_N} = e^{i\pi/2} \Ket{q_j} \;;
\label{eqtensor}
\end{equation}
see Ref.~\cite{Schack2000a} for the reason for the phase factor $e^{i\pi/2}$.
Momentum and position eigenstates are related through the quantum Fourier
transform operator $\hat F$ {\cite{Saraceno1990}}, i.e., 
$\hat F\Ket{q_k}=\Ket{p_k}$.

By applying the Fourier transform operator to the $n$ rightmost bits of
the position eigenstate $|.\xi_{n+1}\ldots\xi_N\xi_n\ldots\xi_1\rangle$, 
one obtains the family of states  \cite{Schack2000a}
\begin{eqnarray}                                                 \label{baker8}
|\xi_1\ldots\xi_n.\xi_{n+1}\ldots\xi_N\rangle&\equiv&
2^{-n/2}e^{i\pi(0.\xi_n\ldots\xi_11)}
|\xi_{n+1}\rangle
            \otimes\cdots\otimes
            |\xi_N\rangle \otimes\cr
    & & (|0\rangle+e^{2\pi i(0.\xi_11)}|1\rangle)
                     \otimes(|0\rangle+e^{2\pi i(0.\xi_2\xi_11)}|1\rangle)
                     \otimes\cr
    & & (|0\rangle+e^{2\pi i(0.\xi_3\xi_2\xi_11)}|1\rangle)
                      \otimes\cdots\otimes\cr
    & & (|0\rangle+e^{2\pi i(0.\xi_n\ldots\xi_11)}|1\rangle)\;,
\end{eqnarray}
where $1\leq n\leq N-1$. 
For fixed values of $n$ and $N$ we will use the notation 
\begin{equation} \label{Eq:basis_n}
|\xi_1\ldots\xi_N\rangle_n\equiv
|\xi_1\ldots\xi_n.\xi_{n+1}\ldots\xi_N\rangle\,.
\end{equation}
These states form an orthonormal basis of the Hilbert space.  
The state (\ref{baker8}) is localised in both 
position and momentum: it is strictly localised within a position region 
of width $1/2^{N-n}$, centred at position 
$q=0.\xi_{n+1}\ldots\xi_N1$, and it is approximately localised within 
a momentum region of width $1/2^{n}$, centred at momentum 
$p=0.\xi_n\ldots\xi_11$.

For each fixed $n$, $0\leq n\leq N-1$, the quantum
baker's map $B_n$ is defined by
\begin{equation}                                                 \label{baker9}
B_n|\xi_1\ldots\xi_n.\xi_{n+1}\ldots\xi_N\rangle =
                    |\xi_1\ldots\xi_{n+1}.\xi_{n+2}\ldots\xi_N\rangle \;,
\end{equation}
i.e.
\begin{equation}                                                \label{baker10}
B_n|\xi_1\ldots \xi_N\rangle_n =|\xi_1\ldots\xi_N\rangle_{n+1}\;.
\end{equation}
\noindent
The action of the map $B_n$ on the basis states
(\ref{baker8}) is thus given by a shift of the dot by one
 position. In phase-space language, the map $\hat B_n$ takes a state 
localised at
$(q,p)=(0.\xi_{n+1}\ldots\xi_N1,0.\xi_n\ldots\xi_11)$ to a state localised at
$(q',p')=(0.\xi_{n+2}\ldots\xi_N1,0.\xi_{n+1}\ldots\xi_11)$, 
while it stretches the
state by a factor of two in the $q$ direction and squeezes it by a
factor of two in the $p$ direction.
For $n=N-1$, the map is the original quantum baker's map
as defined in Ref.~\cite{Saraceno1990}.

For the sake of clarity, it will be convenient to simplify our
notation slightly. Throughout the paper $n$ and $N$ are fixed. 
So we may omit the index $n$ and denote the quantum baker's map 
simply by $B$, always keeping in mind that we are dealing 
with the special baker's map $B_n$ for the given value of $n$.

\subsection{Decoherent histories formalism} \label{sec:DecHist}

The decoherent histories
formalism~\cite{Gell-MannHartle1993,Griffiths1984,Omnes1988,
  Gell-MannHartle1990,DowkerHalliwell1992} provides a framework for
investigating classicality in quantum
theory~\cite{DowkerHalliwell1992,Gell-MannHartle1993}.
The formalism assigns probabilities to quantum histories, 
i.e.\ ordered sequences of quantum-mechanical ``propositions''. 
Mathematically, these propositions are represented by projectors. 
An exhaustive set of mutually exclusive propositions
corresponds to a complete set of mutually orthogonal projectors.
In this approach to quantum theory a quantum mechanical system is 
said to exhibit classical behaviour when the probability distribution 
over histories is strongly peaked about histories
having correlations in time implied by classical deterministic 
laws~\cite{Gell-MannHartle1993,BrunHartle1999-PRD}.  Due to quantum 
interference one cannot always assign probabilities 
to a set of histories in a consistent way. For this to be possible, 
the set of histories must be decoherent. Decoherence of histories 
is therefore a prerequisite for classical behaviour. In general, only
coarse-grained sets of histories are decoherent.

For our purpose it will be sufficient to consider a slightly 
simplified version of the general decoherent histories 
framework, tailored  to a system dynamics 
induced by a fixed unitary quantum map $U$ and restricted to 
the special but natural case, in which histories are constructed 
from a fixed exhaustive set of mutually exclusive propositions.

A {\em projective partition\/} of a Hilbert space $\cal H$ is
a complete set of mutually orthogonal projection operators $\,\{P_{\mu}\}$ on
${\cal H}$, i.e., $P_{\mu}P_{\mu'}=\delta_{\mu\mu'}P_{\mu}\:$ and 
$\:\sum_{\mu}P_{\mu}=\one_{\cal H}\:$, where $\one_{\cal H}$ denotes
the unit operator on $\cal H$. 
A projective partition is {\em
  fine-grained\/} if all projectors are one-dimensional, 
i.e.,  $\,\forall\,\mu\;$
$\mbox{rank}(P_{\mu})=\mbox{dim}\big(\mbox{supp}(P_{\mu})\big)=1$
\footnote{The support of a Hermitian operator $A$ is defined to 
be the vector space spanned by the eigenvectors of $A$ corresponding 
to its non-zero eigenvalues.}, and {\em coarse-grained} otherwise. 

Given a projective partition $\{P_{\mu}\}$ of a Hilbert space $\cal H$, 
a string of length $k$ of projectors $P_{\alpha}\in\{P_{\mu}\}$ 
defines a history of length $k$:
\begin{equation}
h_{\balphas} \equiv \left(P_{\alpha_1},P_{\alpha_2},\dots,P_{\alpha_k}\right)\;,
\end{equation}
where $\balpha\equiv\alpha_1\alpha_2\dots\alpha_k$. The set of all 
such histories, $\mathbb{H}[\{P_{\mu}\}\,;\,k\,]\equiv\big\{h_{\balphas}\,:\:
h_{\balphas}\in\{P_{\mu}\}^k\big\}$, forms the 
exhaustive set of mutually exclusive histories of length
$k$. Histories are ordered sequences of projection
operators, corresponding to quantum-mechanical propositions. 
Note that we restrict attention to histories constructed from a 
fixed exhaustive set of mutually exclusive propositions: 
the projectors $P_{\alpha_{j}}$ within the 
sequences are all chosen from the same projective partition, 
for all times $j=1,\ldots,k$. 

A set of histories  $\mathbb{H}[\{P_{\mu}\}\,;\,k\,]$ 
is called {\em fine-grained\/} ({\em coarse-grained\/}) if it is 
constructed from a fine-grained (coarse-grained) projective partition.
A single history $h_{\balphas}\in\mathbb{H}[\{P_{\mu}\}\,;\,k\,]$ 
is called {\em fine-grained}, if it is represented by 
a sequence of 1-dimensional projectors, and {\em coarse-grained\/} otherwise.

An initial state represented by a density operator $\rho_0$ on ${\cal H}$ and
a unitary dynamics generated by a unitary map $U:\cal H\rightarrow\cal H$
induce a probabilistic structure on the event algebra associated with
$\mathbb{H}[\{P_{\mu}\}\,;\,k\,]$, if the following decoherence conditions are
satisfied.  These are given in terms of properties of the {\em decoherence
  functional\/} $\mathcal{D}_{U,\,\rho_0}\,[\cdot,\cdot]$ on
$\mathbb{H}[\{P_{\mu}\}\,;\,k\,]\times\mathbb{H}[\{P_{\mu}\}\,;\,k\,]$,
defined by
\begin{equation} 
\mathcal{D}_{U,\,\rho_0}\,[h_{\balphas},h_{\bbetas}]\equiv 
\mbox{Tr}\left[C_{\balphas}\,\rho\,C_{\bbetas}^{\dagger}\right]\:,
\end{equation}
where 
\begin{eqnarray} \label{Classoperators}
C_{\balphas}\equiv C_{h_{\mbox{\tiny{$\balpha$}}}}
& \equiv &\left(U^{\dagger\,k}P_{\alpha_k}U^k\right)
\left(U^{\dagger\,k-1}P_{\alpha_{k-1}}U^{k-1}\right)\dots
 \left(U^{\dagger}P_{\alpha_1}U\right)\nonumber\\
&=& U^{\dagger\,k}P_{\alpha_k}UP_{\alpha_{k-1}}U\dots
  P_{\alpha_2}UP_{\alpha_1}U\;.
\end{eqnarray}
The set of histories $\mathbb{H}[\{P_{\mu}\}\,;\,k\,]$ is said to be 
decoherent with respect to a given 
unitary map $U:\cal H\rightarrow\cal H$ and a given initial 
state $\rho_0$, if 
\begin{equation}  \label{eq:consistency}
\mathcal{D}_{U,\,\rho_0}\,[h_{\balphas},h_{\bbetas}]
\propto \delta_{\balphas\bbetas}\equiv
\prod_{j=1}^k\delta_{\alpha_j \beta_j}
\end{equation}
for all $h_{\balphas},h_{\bbetas}\in
\mathbb{H}[\{P_{\mu}\}\,;\,k\,]$. 
If this decoherence condition is satisfied, the diagonal elements 
of the decoherence functional, 
$p[h_{\balphas}]=\mathcal{D}_{U,\,\rho_0}\,[h_{\balphas},h_{\balphas}]$,
can be interpreted as the probabilities of the histories.
For a decoherent set of histories, the entropy, $H[\{h_{\balphas}\}]$,
can be defined as follows \cite{Gell-MannHartle1990,Hartle1998,BrunHartle1999-PRE}:
\begin{eqnarray}        \label{def:Entropy}
H[\{h_{\balphas} \}] 
 &\equiv& -\sum_{\balphas}p[h_{\balphas}]\log_2p[h_{\balphas}]\nonumber\\
&=&-\sum_{\balphas}\mathcal{D}_{U,\,\rho_0}\,[h_{\balphas},h_{\balphas}]
\log_2\Big(\mathcal{D}_{U,\,\rho_0}\,[h_{\balphas},h_{\balphas}]\Big)\;.
\end{eqnarray}

\section{Predictability for different coarse-grainings of the quantum baker's map } 
\label{sec:coarse}

This section is organised as follows.
Subsection~\ref{sec:DifferentCoarse-grainedDescriptions}, which discusses
coarse-grained descriptions of the quantum baker's map, contains two parts:
part~1 introduces a family of coarse-grained projective partitions of the
Hilbert space, which are then used in part~2 to construct a class of
coarse-grained sets of histories.  Subsection~\ref{sec:Results}
summarises the main results of this paper, which are then derived and
illustrated in detail in Subsection~\ref{sec:Derivations}.

\vspace{-2mm}
\subsection{Coarse-grainings}
\label{sec:DifferentCoarse-grainedDescriptions}
\vspace{-1mm}

\subsubsection{Coarse-grained partitions}
\label{sec:Coarse-grained_partitions}

Let us first introduce two different types of coarse-grained 
projective partitions of the $2^N$-dimensional Hilbert space modelling 
the unit square, which later will be regarded as special cases
of a family of more general coarse-grained descriptions. 
We refer to the definitions and notations of Sec.~\ref{sec:baker}. 
In particular we use the orthonormal basis (\ref{Eq:basis_n}) of 
the Hilbert space to construct the partitions. 

For a fixed binary string $\y=y_1\ldots y_{N-l-r}\in \{0,1\}^{N-l-r}$ 
we define the ``local'' projection operators by 
\begin{equation}                     \label{coarseProjectorsloc}
  {P}_{\ys}^{(l,r)}\equiv
\sum_{a_1,\ldots,a_l\atop b_1,\ldots,b_r}
|a_1\dots a_l\ \y\ b_1\dots b_r\rangle_n\,_n
\langle a_1\dots a_l\ \y\ b_1\dots b_r|\,
\equiv\sum_{\as\in \{0,1\}^{l} \atop \bs\in \{0,1\}^{r}} 
|\at\ \y\ \bt\ \rangle_n\,_n
\langle \at\ \y\ \bt|\,,
\end{equation} 
\noindent
and for fixed strings $\y^1\in \{0,1\}^{s_1}$ and $\y^2\in
\{0,1\}^{s_2}$ we define the ``nonlocal'' 
projection operators by
\begin{eqnarray}                     \label{CheckerboardProjectors}
  {P}_{\ys^1,\,\ys^2}^{(l,m_l, m_r, r)}&\equiv &
\sum_{\as\in \{0,1\}^{l}}\:\sum_{\bs\in \{0,1\}^{r}}\:
\sum_{\,\xibolds\in \{0,1\}^{m_l+m_r}}|\at\ \y^1\ \xibold\ \y^2\ \bt\ \rangle_n\,_n
\langle\at\ \y^1\ \xibold\ \y^2\ \bt\ |\;\\
 &\equiv&\sum_{a_1,\ldots,a_l\atop b_1,\ldots,b_r}
\sum_{\xi_1\ldots\xi_{m_l}\atop \xi_{m_l+1}\ldots\xi_{m_l+m_r}}
|a_1\dots a_l\ \y^1\ \xi_1\ldots\xi_{m_l}\,.\,\xi_{m_l+1}\ldots\xi_{m_l+m_r}
 \y^2\ b_1\dots b_r\rangle\times\nonumber\\ && \hspace{3.5cm}\langle a_1\dots a_l\ \y^1\
\xi_1\ldots\xi_{m_l}\,.\,\xi_{m_l+1}\ldots\xi_{m_l+m_r}  \y^2\ b_1\dots
b_r|\nonumber
\end{eqnarray}
What the terms ``local'' and  ``non-local'' mean 
in this context, will be explained below. 
Throughout this paper, bold variables denote binary strings.  
Furthermore, lower indices label individual bits of a
string, whereas upper indices will label different strings. 
It will be convenient to abbreviate a substring
$\alpha_{\kappa}\dots\alpha_{\sigma}$ of a string 
$\balpha=\alpha_1\dots\alpha_{\kappa}\alpha_{\kappa+1}\dots\alpha_{\sigma}
\alpha_{\sigma+1}\dots\alpha_{\gamma}$ by $\balpha_{\kappa:\sigma}$. 
Concatenation of strings is defined in the usual way. 
Taking the just mentioned example we can, for instance, express 
the string $\balpha$ as a concatenation of three substrings, 
$\balpha=\balpha_{1:\kappa-1}\balpha_{\kappa:\sigma}\balpha_{\sigma+1:\gamma}$. 
The length of a string $\balpha$ will be denoted by $|\balpha|$. 

For simplicity, we will always assume in the following 
that $l<n$ and $r<N-n$ in the first case, and 
$l+s_1\le n$ and $r+s_2\le N-n$ in the second case. 
In both cases $l$ and $r$ acquire the specific
meaning as the number of ``momentum'' and ``position'' bits 
ignored in the coarse-graining. In the second case, 
in addition $m_l$ most significant momentum 
bits and $m_r$ most significant position 
bits are coarse-grained over. 

The operator ${P}_{\ys}^{(l,r)}$ is a projector on a
$2^{l+r}$-dimensional subspace labelled by the string $\y$. 
The projector ${P}_{\ys^1,\,\ys^2}^{(l,m_l, m_r, r)}$ projects 
on a $2^{l+m_l+m_r+r}$-dimensional subspace labelled by the pair 
of strings $(\y^1, \y^2)$. 
In both cases we are dealing with complete sets of
mutually orthogonal projectors, i.e., with projective partitions, as
\begin{equation}
{P}_{\ys}^{(l,r)}{P}_{\ys'}^{(l,r)}=\delta_{\ys,\ys'}{P}_{\ys}^{(l,r)}\quad 
\mbox{and}\quad\sum_{\ys}{P}_{\ys}^{(l,r)}=\one\;,
\end{equation}
\begin{equation}
{P}_{\ys^1,\,\ys^2}^{(l,m_l, m_r, r)}{P}_{\ys^{'1},\,\ys^{'2}}^{(l,m_l, m_r, r)}=
\delta_{\ys^1,\,\ys^{'1}}\delta_{\ys^2,\,\ys^{'2}}
{P}_{\ys^1,\,\ys^2}^{(l,m_l, m_r,r)}\quad\mbox{and}\quad
\sum_{\ys^1,\,\ys^2}{P}_{\ys^1,\,\ys^2}^{(l,m_l, m_r, r)}=\one\;.
\end{equation}

\begin{figure}[bt]
\begin{center}
\includegraphics[width=10cm]{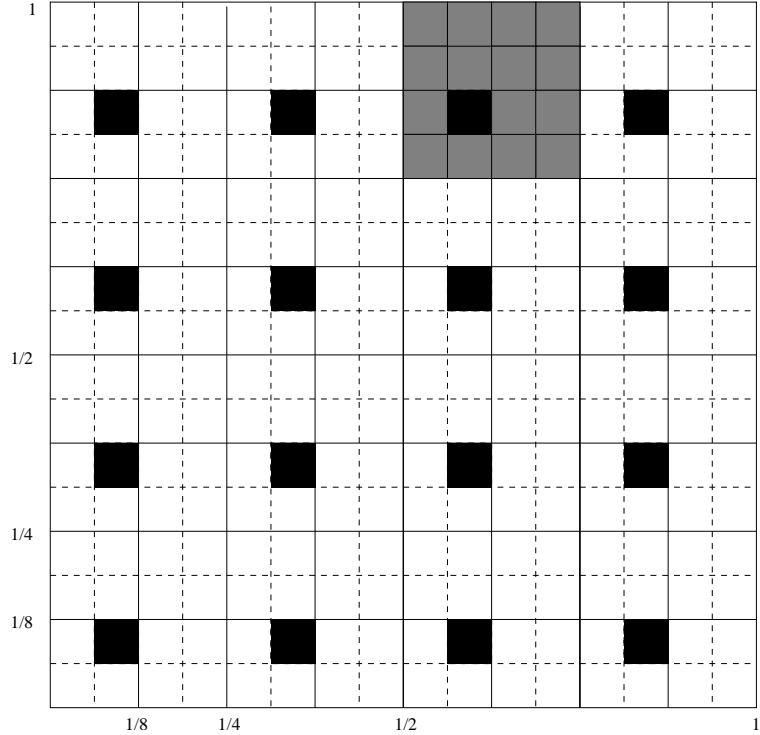}
\bigskip
\bigskip
\caption{A schematic illustration of the projectors  ${P}_{\ys}^{(l,r)}$ and
${P}_{\ys^1,\,\ys^2}^{(l,m_l, m_r, r)}$. {\bf (a)} Let 
  $n-l=2$, $N-l-r=4$, and $\y=y_1\ldots y_4=1110$. The projector
  ${P}_{\ys=1110}^{(l,r)}$ is then in approximate correspondence with
  the phase space region shaded in grey. {\bf (b)} Let $n-l=4$, $m_l=2$,
  $m_r=2$, $\y^1=10$ and $\y^2=01$. The projector
  ${P}_{\ys^1=10,\,\ys^2=01}^{(l,2,2, r)}$ is then in approximate
  correspondence with the disconnected phase space region given by the 16
  black cells.}
\label{fig1}
\end{center}
\end{figure}

Let us explain what is meant  by ``local'' and ``nonlocal'' 
regarding the just introduced projection operators.  
The projection operators ${P}_{\ys}^{(l,r)}$ and
${P}_{\ys^1,\,\ys^2}^{(l,m_l, m_r, r)}$ 
project on subspaces of the Hilbert space associated 
with phase-space regions of the unit square in which 
the corresponding eigenstates with eigenvalue 1  are localised. 
In the case of the projectors ${P}_{\ys}^{(l,r)}$  
these regions are connected cells whose location 
within the unit square of the phase space is determined 
by the specified most significant position and momentum 
bits given by the binary string $\y=y_1\ldots y_{N-l-r}$. 
The size of these cells depends on the significance of 
the scales which are not resolved and therefore ignored, i.e.\ coarse-grained over.   
In the case of the projectors ${P}_{\ys^1,\,\ys^2}^{(l,m_l, m_r, r)}$, 
on the other hand, there is coarse-graining also at 
the most significant scales:  a number of the most significant
position and  momentum bits are not specified.  The  associated phase space domains
must therefore consist of disconnected parts spread over the whole 
unit square, the number depending on how many most significant
position and momentum bits are coarse-grained over, i.e.\ on the 
parameter $m\equiv m_l+m_r$.  For an illustration see Fig.~\ref{fig1}. 

We will also use the diagram notation for the introduced projectors: 
\begin{eqnarray} 
\label{coarseProjectorsloc_diagram}
{P}_{\ys}^{(l,r)}&\equiv&(\,\underbrace{\Box\Box\dots\Box}_{l}
\ \y\ \underbrace{\Box\Box\dots \Box}_{r}\,)\;,\\
\label{CheckerboardProjectors_diagram}
 {P}_{\ys^1,\,\ys^2}^{(l,m_l, m_r, r)}&\equiv& (\,\underbrace{\Box\Box\dots\Box}_{l}
\ \y^1\ \underbrace{\Box\Box\dots\Box}_{m_l}\ .\ \underbrace{\Box\Box\dots \Box}_{m_r}\
\y^2\ \underbrace{\Box\Box\dots \Box}_{r}\,)\;,
\end{eqnarray}
where the empty boxes indicate the bits which are coarse-grained
over. We can write the projectors of the second type as 
sums over projectors of the first type: 
\begin{equation}    \label{nonlocal=sumoverlocal}
{P}_{\ys^1,\,\ys^2}^{(l,m_l, m_r, r)}= \sum_{\,\xibolds\in \{0,1\}^{m_l+m_r}}  
{P}_{\ys^1\xibolds\ys^2}^{(l,r)}\;,
\end{equation}
where $\y^1\xibold\y^2$ means the concatenation of the three strings
$\y^1$, $\xibold$ and $\y^2$. Remember that in the definition of 
the projectors ${P}_{\ys^1,\,\ys^2}^{(l,m_l, m_r, r)}$ 
we assume that $l+|\y^1|\le n$ and 
$r+|\y^2|\le N-n$. 

The projectors (\ref{coarseProjectorsloc_diagram}) and 
(\ref{CheckerboardProjectors_diagram}) are special cases 
of the family of all projection operators, which define the 
scales at which information is lost in the symbolic 
representation. In general such projectors exhibit 
structure on many different scales, and the most 
general projector of this type would be of the form
\begin{equation}
\label{multi-scales_coarse-graining_diagram}
{P}_{\ys^1,\,\ys^2,\dots,\ys^{\lambda}}^{(l,m_1, m_2,\dots,m_{\lambda-1},r)}
=
\bigg( \,\underbrace{\Box\dots\Box}_{l}
\ \y^{1}\ \underbrace{\Box\dots\Box}_{m_1}\,\y^{2}\ 
\underbrace{\Box\dots\Box}_{m_2}\ \,\dots\dots\,
\y^{\lambda-1}\ \underbrace{\Box\dots\Box}_{m_{\lambda-1}}\
\y^{\lambda}\  \underbrace{\Box\dots\Box}_{r}\,\bigg)\;. 
\end{equation}
The projector (\ref{multi-scales_coarse-graining_diagram}) 
defines a coarse-graining in which information is lost on 
several different scales. We will call this a {\em multi-scale
  coarse-graining\/} or {\em hierarchical coarse-graining}.
Accordingly, the special cases  
(\ref{coarseProjectorsloc_diagram}) and 
(\ref{CheckerboardProjectors_diagram}) will be called 
1-scale and 2-scale coarse-graining, respectively. 
The  2-scale coarse-graining (\ref{CheckerboardProjectors_diagram}) 
we introduced above is a special 2-scale coarse-graining, 
as we assumed that the coarse-grained {\em island\/} of 
size $m_l+m_r$ between the specified strings $\y^1$ and 
 $\y^2$ lies around the dot separating the momentum and 
position bits in the symbolic representation. The first step 
towards a generalisation is to combine the two parameters 
$m_l$ and $m_r$ (i.e.\ the number of 
most significant momentum and position bits that 
are coarse-grained over in the symbolic representation)  
to a single parameter $m=m_l+m_r$ and allow the corresponding 
coarse-grained island of size $m$ between the specified strings $\y^1$ and 
 $\y^2$ to lie anywhere, not necessarily at 
the most significant region around the dot. 
The next step is to introduce several 
coarse-grained islands of this kind, on several scales.
An event will then be specified by bit strings $\y^1$, $\y^2$, 
$\dots$, $\y^{\lambda}$ of length $|\y^i|=s_i$ at a time, separated by 
$(\lambda-1)$  coarse-grained islands of size $m_i$ each, where $\lambda>1$,
as in Eq.~(\ref{multi-scales_coarse-graining_diagram}). 

More precisely, the most general family of coarse-grained descriptions is  
represented by sets of projection operators defined as follows:
\begin{eqnarray}                     \label{multi-scales_coarse-graining_projectors}
  {P}_{\ys^1,\,\ys^2,\dots,\ys^{\lambda}}^{(l,m_1, m_2,\dots,m_{\lambda-1},r)}&\equiv &
\sum_{\as\in \{0,1\}^{l}}\:\sum_{\bs\in \{0,1\}^{r}}\:
\sum_{\,\xibolds^1\in \{0,1\}^{m_1}}\dots
\sum_{\,\xibolds^{\lambda-1}\in \{0,1\}^{m_{\lambda-1}}}\nonumber\\
&&|\at\ \y^1\ \xibold^1\ \y^2\ \xibold^2\ \dots \xibold^{\lambda-1}
\y^{\lambda}\bt\ \rangle_n\,_n\langle
\at\ \y^1\ \xibold^1\ \y^2\ \xibold^2\ \dots
\xibold^{\lambda-1}\y^{\lambda}\bt\ |\nonumber\\
&=&
\sum_{\,\xibolds^1\in \{0,1\}^{m_1}}\dots
\sum_{\,\xibolds^{\lambda-1}\in \{0,1\}^{m_{\lambda-1}}}
{P}_{\ys^1\xibolds^1\ys^2\xibolds^2\dots
\ys^{\lambda-1}\xibolds^{\lambda-1}\ys^{\lambda}}^{(l,r)}\;,
\end{eqnarray}
\noindent
where $\y^1\xibold^1\y^2\xibold^2\dots
\y^{\lambda-1}\xibold^{\lambda-1}\y^{\lambda}$ 
means the concatenation of the particular strings 
$\y^1$, $\xibold^1$, $\y^2,\dots,\xibold^{\lambda-1}$,
$\y^{\lambda}$. We still assume $l<n$ and $r<N-n\,$. 
Eq.~(\ref{multi-scales_coarse-graining_diagram}) 
is a diagram notation of
Eq.~(\ref{multi-scales_coarse-graining_projectors}). 
It is easily seen that for fixed $m_1,\dots,m_{\lambda-1}$  the 
set $\;\{{P}_{\ys^1,\,\ys^2\,,\dots ,\, \ys^{\lambda}}^{(l,m_1,
  m_2\,,\dots ,\,m_{\lambda-1}, r)}\}$ forms 
a projective partition of the Hilbert space, as
\begin{eqnarray}
{P}_{\ys^1,\,\ys^2\,,\dots ,\, \ys^{\lambda}}^{(l,m_1,
  m_2\,,\dots ,\,m_{\lambda-1}, r)}
{P}_{\ys^{'1},\,\ys^{'2}\,,\dots ,\, \ys^{'\lambda}}^{(l,m_1,
  m_2\,,\dots ,\,m_{\lambda-1}, r)}&=&
\delta_{\ys^1,\,\ys^{'1}}\delta_{\ys^2,\,\ys^{'2}}\times\dots
\times\delta_{\ys^{\lambda},\,\ys^{'\lambda}}
{P}_{\ys^1,\,\ys^2\,,\dots ,\,\ys^{\lambda}}^{(l,m_1,
  m_2\,,\dots ,\,m_{\lambda-1}, r)}\nonumber\\
\quad\mbox{and}\quad
\sum_{\ys^1,\,\ys^2,\dots, \ys^{\lambda}}
{P}_{\ys^1,\,\ys^2\,,\dots ,\, \ys^{\lambda}}^{(l,m_1,
  m_2\,,\dots ,\,m_{\lambda-1}, r)}&=&\one\;.
\end{eqnarray}

\subsubsection{Coarse-grained histories}

In order to investigate coarse-grained  evolution we now construct 
coarse-grained  histories. By considering different types of histories 
constructed from different types of coarse-grained projective partitions 
we obtain different coarse-grained effective evolutions.
Our investigation of the coarse-grained  evolution of the 
quantum baker's map starts with 
the special cases of 1-scale and 2-scale coarse-grainings as
defined in Eqs.~(\ref{coarseProjectorsloc_diagram}) and 
(\ref{CheckerboardProjectors_diagram}). We first  compare
the different members of the family 
\begin{eqnarray}
\label{family_ofcoarse-grainings:set}
\bigg\{\left\{{P}_{\ys^1,\,\ys^2}^{(l,m_l, m_r, r)}\,:\,
\y^1\in \{0,1\}^{s_1}\,,\,\y^2\in
\{0,1\}^{s_2}\right\}\,&:&\:l,r,m_l,m_r,s_1,s_2\in
\{0,1,2,\dots\}\;\;\phantom{\bigg\}}\nonumber\\
\phantom{\bigg\{}\mbox{such that}\quad l+s_1\le n\,,\,r+s_2\le N-n
&\mbox{and}&
l+r+s_1+s_2+ m_l+m_r=N\bigg\}\nonumber\\&&
\end{eqnarray}
of coarse-grained descriptions, parameterised by 
$l$, $r$, $s_1$, $s_2$,  $m_l$ and $m_r$,   
with respect to predictability of the evolution. 
Our results will concern only such members of this family
for which $s_1$ and $s_2$ are significantly greater than $1$, and $s_1\ge m_l+m_r$. 
Furthermore, in order to obtain the classical limit of the quantum 
baker's map, we will be considering only members with very large 
value for the parameter $l$, as $\hbar\rightarrow 0$ will correspond 
to $l \rightarrow \infty$. Finally, the results will show that only 
$m=m_l+m_r$ matters, and the specification ``$m_l$ most 
significant momentum bits and $m_r$ most significant position bits are 
coarse-grained over'' therefore  be unnecessary. Note that the local 1-scale 
coarse-graining (\ref{coarseProjectorsloc}) is included  in 
this family as the special case $m_l+m_r=0$. 

The histories corresponding to 1-scale and 2-scale coarse-graining 
(\ref{coarseProjectorsloc_diagram}) and 
(\ref{CheckerboardProjectors_diagram}) will be 
labelled by finite sequences of strings in the first case and pairs of  
finite sequences of strings in the second case, respectively:
\begin{eqnarray}                                      \label{LocalHistories}
h_{\vec{\ys}}&\equiv&
                \big({P}^{(l,r)}_{\ys^1},{P}^{(l,r)}_{\ys^2},
\dots,{P}^{(l,r)}_{\ys^k}\big)_{\phantom{|_{|_{|_{|_|}}}}}\;, 
\end{eqnarray}
where $\vec{\y}= (\y^1, \dots ,\y^k)$ is a sequence of strings
$\y^j\in\{0,1\}^{N-l-r}$, $j=1, \dots , k$;
\begin{eqnarray}                                      \label{CheckerboardHistories}
h_{\vec{\ys}^1,\,\vec{\ys}^2}&\equiv&
                \big({P}_{\ys^{1,1},\,\ys^{1,2}}^{(l,m_l, m_r, r)}, 
                     {P}_{\ys^{2,1},\,\ys^{2,2}}^{(l,m_l, m_r, r)}, \dots,
                     {P}_{\ys^{k,1},\,\ys^{k,2}}^{(l,m_l, m_r,
                r)}\big)_{\phantom{|_{|_{|_{|_|}}}}}\;,  
\end{eqnarray}
where $(\vec{\y}^1,\vec{\y}^2) = ((\y^{1,1}, \dots ,\y^{k,1}),(\y^{1,2}, \dots ,\y^{k,2}))$
is a pair of finite sequences of strings $\y^{j,i}\in\{0,1\}^{s_i}$,
$j=1, \dots , k$, $i=1,2$, labelling the history.

To examine decoherence of this set of histories and calculate its probability
distribution we will evaluate  the decoherence functional
\begin{equation}                                      \label{dfunc_local}
{\mathcal D}_{B,\,\rho_0}[h_{\vec{\ys}},
                                    h_{\vec{\zs}}   ]
=
                \Tr[P^{(l,r)}_{\ys^k}B P^{(l,r)}_{\ys^{k-1}}
                B\cdots P^{(l,r)}_{\ys^1}
B\rho_0 B^\dag P^{(l,r)}_{\zs^1}
   \cdots
         B^\dag P^{(l,r)}_{\zs^{k-1}}B^\dag P^{(l,r)}_{\zs^k}] \;,
\end{equation}
and           
 \begin{equation*}                           \label{checkerboard-dfunc}  
{\mathcal D}_{B,\,\rho_0}[h_{\vec{\ys}^1,\,\vec{\ys}^2},
  h_{\vec{\zs}^1,\,\vec{\zs}^2} ]
\equiv \hspace{11cm}
\end{equation*}
 \begin{equation}  \label{checkerboard-dfunc}  
               = \Tr[{P}_{\ys^{k,1},\,\ys^{k,2}}^{(l,m_l, m_r, r)}B
                    {P}_{\ys^{k-1,1},\,\ys^{k-1,2}}^{(l,m_l, m_r, r)} B\cdots
                    {P}_{\ys^{1,1},\,\ys^{1,2}}^{(l,m_l, m_r, r)}
                    B\rho_0 B^\dag
                    {P}_{\zs^{1,1},\,\zs^{1,2}}^{(l,m_l,m_r,r)}B^\dag\cdots
                    B^\dag
                    {P}_{\zs^{k,1},\,\zs^{k,2}}^{(l,m_l, m_r, r)}] \;,
\end{equation}
respectively.

Whether the decoherence functional is diagonal or not depends
on the initial state $\rho_0$. In order to check decoherence 
of a given set of histories and assign probabilities 
to them  we therefore need to specify the initial state from which 
the histories start. 
Here we choose a certain class of states as the initial states 
for the histories, namely the discrete set of states that are 
induced via normalisation by the set of projectors defining the histories. 
We therefore assume the initial state $\rho_0$ to be of the same form 
as the events in the histories, i.e.\ to be proportional to one of 
the projection operators of the set $\{  {P}_{\ys}^{(l,r)}\}$ or 
$\{  {P}_{\ys^1,\,\ys^2}^{(l,m_l, m_r, r)}\}$, respectively:  
\begin{equation}        
\label{LocalInitialstate}               
\rho_0 = \rho_{\xs}^{(l, r)}\equiv
2^{-(l+r)} {P}_{\xs}^{(l,r)}\\
\equiv 2^{-(l+r)} (\,\underbrace{\Box\Box\dots\Box}_{l}
\ \x\ \underbrace{\Box\Box\dots \Box}_{r}\,)\;,
\end{equation}
or
\begin{eqnarray}        \label{CheckerboardInitialstate}               
\rho_0 &=& \rho_{\xs^1,\,\xs^2}^{(l,m_l, m_r, r)}\equiv
2^{-(l+m_l+m_r+r)} {P}_{\xs^1,\,\xs^2}^{(l,m_l, m_r, r)}\\
&\equiv& 2^{-(l+m_l+m_r+r)} (\,\underbrace{\Box\Box\dots\Box}_{l}
\ \x^1\ \underbrace{\Box\Box\dots\Box}_{m_l}\ .\ 
\underbrace{\Box\Box\dots \Box}_{m_r}\
\x^2\ \underbrace{\Box\Box\dots \Box}_{r}\,)\;.\nonumber
\end{eqnarray}
The normalisation factor $2^{-(l+r)}$ or $2^{-(l+m_l+m_r+r)}$, 
respectively,  ensures that 
$\rho_0$ is a density operator, i.e.\ $\Tr[\rho_0]=1$. 
All calculations in Sec.~\ref{sec:Derivations} will be based 
on this choice for the initial states, which we regard 
as the most natural choice within our framework of sets of 
histories constructed from a given, fixed projective 
partition.   

We now generalise the family of sets of coarse-grained 
histories from the 1-scale and 2-scale coarse-grained 
descriptions considered above to the general case 
of multi-scale (or hierarchical) coarse-grainings. The corresponding 
projective partitions have already been introduced 
in Eqs.~(\ref{multi-scales_coarse-graining_diagram}) 
and (\ref{multi-scales_coarse-graining_projectors}). 
The generalised family of coarse-grained descriptions 
is therefore given by the set:
\begin{eqnarray}
\label{family_hierarchical_coarse_grainings}
\bigg\{\left\{{P}_{\ys^1,\,\ys^2\,,\dots ,\, \ys^{\lambda}}^{(l,m_1,
  m_2\,,\dots ,\,m_{\lambda-1}, r)}\,\right\}_{\y^j\in \{0,1\}^{s_j}}\,&:&\:\:
l,r,m_j, s_j\in\{0,1,2,\dots\}\phantom{\bigg\}}
\,,\;\lambda\in\{1,2,3,\dots\}\nonumber\\
\phantom{\bigg\{}\mbox{such that}
&&l+r+\sum_{j=1}^{\lambda-1} m_j+\sum_{j=1}^{\lambda}s_j=N
\bigg\} \;.
\end{eqnarray}
The members of this family are represented by coarse-grained projective partitions
displaying coarseness on several different scales in the symbolic
representation.  The family is parameterised by $l,r, m_1,\dots
,\,m_{\lambda-1}$, $s_1,\dots,s_{\lambda}$ and $\lambda$   
with the  constraint 
$l+r+\sum_{j=1}^{\lambda-1} m_j+\sum_{j=1}^{\lambda}s_j=N$.  
Again, our results will involve only such members of this family,  
for which $s_1,\dots,s_{\lambda}$ have values significantly greater 
than $1$, and the value of $l$ is very large (classical limit). 

Our generalised type of histories is  labelled by (finite) sequences 
of finite sequences of binary strings: 
\begin{equation}\label{set:hierarchically-coarse-grained-histories}
\bigg\{h_{\vec{\ys}^1,\,\vec{\ys}^2,\dots,\vec{\ys}^{\lambda}}\;:\;\vec{\y}^{\,i}= 
(\y^{1,i}, \dots ,\y^{k,i})\quad\mbox{with}\quad \y^{j,i}\in\{0,1\}^{s_i}\;,\;
j=1,\dots,k\;,\;  i=1,\dots,\lambda\bigg\}\;.
\end{equation}
They are explicitely defined by time-ordered sequences of 
(\ref{multi-scales_coarse-graining_projectors})-type projection 
operators: 
\begin{equation} 
h_{\vec{\ys}^1,\,\vec{\ys}^2,\dots,\vec{\ys}^{\lambda}}
\equiv
\Big({P}_{\ys^{1,1},\,\ys^{1,2},\dots,\ys^{1,\lambda}}
           ^{(l,m_1, m_2,\dots,m_{\lambda-1},r)}\;,\;
       {P}_{\ys^{2,1},\,\ys^{2,2},\dots,\ys^{2,\lambda}}
           ^{(l,m_1, m_2,\dots,m_{\lambda-1},r)}\;,\dots,\;
{P}_{\ys^{k,1},\,\ys^{k,2},\dots,\ys^{k,\lambda}}
           ^{(l,m_1,
  m_2,\dots,m_{\lambda-1},r)}\Big)_{\phantom{|_{|_{|_{|_|}}}}} \;.
\end{equation}

To examine decoherence of the set of 
histories~(\ref{set:hierarchically-coarse-grained-histories}) and calculate
its probability distribution we will evaluate  the decoherence functional            
 \begin{equation*}                     
\hspace*{-10cm}{\mathcal
  D}_{B,\,\rho_0}[h_{\vec{\ys}^1,\,\vec{\ys}^2,\dots,\vec{\ys}^{\lambda}},
h_{\vec{\zs}^1,\,\vec{\zs}^2,\dots,\vec{\zs}^{\lambda}} ]
\equiv
\end{equation*}
\begin{eqnarray}
 \label{hierarchical-dfunc}  
&=\mbox{Tr}\Big[{P}_{\ys^{k,1},\,\ys^{k,2},\dots,\ys^{k,\lambda}}
           ^{(l,m_1,
  m_2,\dots,m_{\lambda-1},r)}B
{P}_{\ys^{k-1,1},\,\ys^{k-1,2},\dots,\ys^{k-1,\lambda}}
           ^{(l,m_1,
  m_2,\dots,m_{\lambda-1},r)}B\cdots
{P}_{\ys^{1,1},\,\ys^{1,2},\dots,\ys^{1,\lambda}}
           ^{(l,m_1,
  m_2,\dots,m_{\lambda-1},r)}B\rho_0 B^\dag\times&\nonumber\\
&\times {P}_{\zs^{1,1},\,\zs^{1,2},\dots,\zs^{1,\lambda}}
           ^{(l,m_1,
  m_2,\dots,m_{\lambda-1},r)}B^\dag\cdots
{P}_{\zs^{k-1,1},\,\zs^{k-1,2},\dots,\zs^{k-1,\lambda}}
           ^{(l,m_1,
  m_2,\dots,m_{\lambda-1},r)}B^\dag
{P}_{\zs^{k,1},\,\zs^{k,2},\dots,\zs^{k,\lambda}}
           ^{(l,m_1,
  m_2,\dots,m_{\lambda-1},r)}\Big] \;,& 
\end{eqnarray}
Again we will choose the initial state to be proportional to one 
of the projection operators defining our coarse-grained 
description, i.e.\ to one of the 
(\ref{multi-scales_coarse-graining_projectors})-type projectors:
\begin{equation}
  \label{Hierarchical_coarse-grainingInitialstate}      
\rho_0 = \rho_{\xs^1,\,\xs^2,\dots,\xs^{\lambda}}^{(l,m_1, m_2,\dots,m_{\lambda-1},r)}\equiv 
2^{-(l+r+m_1+m_2+\dots+m_{\lambda-1})} 
{P}_{\xs^1,\,\xs^2,\dots,\xs^{\lambda}}^{(l,m_1, m_2,\dots,m_{\lambda-1},r)}\;,
\end{equation}
with the normalisation factor ensuring $\Tr[\rho_0]=1$.

\subsection{Main results}
\label{sec:Results}

To characterise and quantify predictability, we use the rate of the entropy
production.  The greater the rate of the entropy production, the more
unpredictable is the evolution. 
We begin by stating the results for the
family~(\ref{family_ofcoarse-grainings:set}) of 1-scale and 2-scale
coarse-grainings.  
First of all we find that in the asymptotic limit $l\to \infty$ 
all the corresponding members of this family 
(i.e., all members with very large parameter value $l$),  
provided that $m_l+m_r$ is finite,  
lead to  decoherent sets of histories, which is the 
prerequisite for classicality. For finite, but very large $l$ 
the decoherence functional is approximately diagonal, which 
means approximate decoherence of histories. 
For very large  $l$, the diagonal elements of the decoherence 
functional,  $\mathcal{D}_{B,\,\rho_0}[h_{\vec{\ys}^1,\,\vec{\ys}^2}, 
h_{\vec{\ys}^1,\,\vec{\ys}^2} ]$, may therefore be interpreted 
as probabilities of the corresponding histories. 
Furthermore we find that for very large $l$,  
for all members of the corresponding subset within this family, 
for which $s_1$ and $s_2$ are significantly greater than $1$,  
the probabilities of the individual alternative histories 
of a set are peaked at histories which display 
regularities according to the {\em classical shift property}. 

We have compared the rates of entropy increase of the different sets 
within the family~(\ref{family_ofcoarse-grainings:set}) of
coarse-grainings. The result for the local coarse-graining 
(\ref{coarseProjectorsloc}), i.e.\ for the case $m_l+m_r=0$, 
was obtained in an earlier work of two of 
us~\cite{Soklakov2002}. In~\cite{Soklakov2002} it was 
shown that in this case the coarse-grained quantum baker's map 
exhibits a linear entropy increase at an asymptotic rate given by the 
Kolmogorov-Sinai entropy~\cite{Alekseev1981} of the classical 
chaotic baker's map, namely 1 bit per iteration step: 
\begin{equation}               \label{EntropyLocal}
H[\{h_{\vec{\ys}} \}] = k+
O(\frac{(l+r-k)\log_2(l+r-k)}{2^{l-2(k^2+k)}})\,, 
\end{equation}
where the set $\{h_{\vec{\ys}} \}$ consists of histories of 
length $k$.

For nonlocal coarse-grainings $m_l+m_r\not=0$, 
the derivation in the next section give these results:
\begin{itemize}
\item Entropy after $k$ iteration steps in case  $k\le m_l+m_r$:
\begin{equation}
H[\{ h_{\vec{\ys}^1,\,\vec{\ys}^2}\}] =
2k\;+\;{\cal
  O}(\frac{(l+r-k)\log_2(l+r-k)}{2^{l-2(k^2+(1+m_l+m_r)k)}})\;
\end{equation}
\item Entropy after $k$ iteration steps in case  $k\ge m_l+m_r$:
\begin{equation}
H[\{ h_{\vec{\ys}^1,\,\vec{\ys}^2}\}] 
=k+\;(m_l+m_r)\;+\;{\cal
  O}(\frac{(l+r-k)\log_2(l+r-k)}{2^{l-2(k^2+(1+m_l+m_r)k)}})\;.
\end{equation}
\end{itemize}

The entropy increase is 2 bits per iteration step as long as the number of
iterations $k$ of the quantum baker's map is smaller than $m=m_l+m_r$.  As
soon as the number of iterations exceeds the parameter $m$, the rate of
entropy increase drops to 1 bit per iteration step. Both short-term and
long-term rates of entropy increase are thus independent of the 
non-locality parameter $m$. The parameter $m$ determines the duration of
the short-term regime for which the entropy increases
at a rate of 2 bits per iterations.

Higher rates of entropy increase become possible for hierarchical
coarse-grainings, i.e.\ coarse-grained histories with coarse-graining on
several different scales of the phase space.  As before we find approximate
decoherence for such sets of histories and a probability distribution which is
peaked at histories displaying regularities according to the classical shift
property. The following results are valid for large $l$
(classical limit) and values for $s_j$ $(j=1,2,\dots,\lambda)$ that are
significantly greater than $1$.
\begin{itemize}
\item Entropy after $k$ iteration steps in the case  
$k<\mbox{min}\{m_1,m_2, \dots, m_{\lambda-1}\}$:
\begin{equation}
H[\{ 
h_{\vec{\ys}^1,\,\vec{\ys}^2,\dots,\vec{\ys}^{\lambda}}\}] 
= 
\lambda\cdot k\;+\;{\cal
  O}(\frac{(l+r-k)\log_2(l+r-k)}{2^{l-2(k^2+(1+m_1+m_2+\dots+m_{\lambda-1})k)}})\;
\end{equation}
\end{itemize}
\begin{itemize}
\item Entropy after $k$ iteration steps in the case  
$k>\mbox{max}\{m_1,m_2, \dots, m_{\lambda-1}\}$:
\begin{equation}
H[\{ 
h_{\vec{\ys}^1,\,\vec{\ys}^2,\dots,\vec{\ys}^{\lambda}}\}] =
k+\sum_{i=1}^{\lambda-1}m_i\;+\;{\cal
  O}(\frac{(l+r-k)\log_2(l+r-k)}{2^{l-2(k^2+(1+m_1+m_2+\dots+m_{\lambda-1})k)}})\;.
\end{equation}
\end{itemize}
We see that in the long-term regime, 
$k>\mbox{max}\{m_1,m_2, \dots, m_{\lambda-1}\}$, 
the rate of entropy increase is again 1 bit per iteration, independently of 
the character of the coarse-graining. In the short-term regime, 
$k<\mbox{min}\{m_1,m_2, \dots, m_{\lambda-1}\}$, however, 
the rate of entropy increase is $\lambda$ bits per iteration. The short-term
regime is thus characterised by $\lambda$,  the number of coarse-graining 
scales. The parameters $m_1,\ldots,m_{\lambda-1}$ determine the duration 
of the  short-term regime. Classical predictability decreases with increasing number
of coarse-graining scales.

Finally, we note how the above results for the entropy
production in the various coarse-grained descriptions can be understood using
the shift property of the coarse-grained evolution of the quantum baker's map,
which is explained and illustrated in detail in the next section. For this we
make use of our diagram notation~(\ref{multi-scales_coarse-graining_diagram}).
The shift property implies that the only histories with significant
probabilities are those that satisfy the shift condition, i.e., the projectors
of the histories have to be related to the initial state via a shift.  For
instance, if $ \rho_0 \propto {P}_{\xs^1,\,\xs^2,\dots,\xs^{\lambda}}^{(l,m_1,
  m_2,\dots,m_{\lambda-1},r)}\;$, then only such histories can arise with
significant probabilities whose first event, represented by the projector
${P}_{\ys^{1,1},\ys^{1,2},\dots, \ys^{1,\lambda}}^{(l,m_1,
  m_2,\dots,m_{\lambda-1},r)}$, satisfies the shift constraint. Unless
$\y^{1,1}_{1:(s_1-1)}=\x^1_{2:s_1}$ and $\y^{1,2}_{1:(s_2-1)}=\x^2_{2:s_2}$
and $\dots$ and $\y^{1,\lambda}_{1:(s_{\lambda}-1)}=\x^\lambda_{2:s_\lambda}$
is satisfied by the first event the whole history will have a vanishing
probability. On the other hand the last bits $y^{1,1}_{s_1}$, $y^{1,2}_{s_2}$,
$\dots$, $y^{1,\lambda}_{s_\lambda}$ of the strings $\y^{1,1}$, $\y^{1,2}$,
$\dots$, $\y^{1,\lambda}$, which denote the first event of the history, remain
undetermined, because the unspecified bits of the empty boxes
in~(\ref{multi-scales_coarse-graining_diagram}), which are coarse-grained
over, are shifted onto them. The bits $y^{1,1}_{s_1}$, $y^{1,2}_{s_2}$,
$\dots$, $y^{1,\lambda}_{s_\lambda}$ may therefore be chosen arbitrarily,
corresponding to a branching into $2^\lambda$ possible histories with
non-vanishing probabilities.  This branching into $2^\lambda$ alternatives
repeats with each iteration step of the evolution, as long as
$k<\mbox{min}\{m_1,m_2, \dots, m_{\lambda-1}\}$, leading to an entropy
production of $\lambda$ bits per iteration step. As soon as the number of
iterations $k$ starts to exceed, step by step, the values of $m_1,m_2, \dots,
m_{\lambda-1}$, the rate of entropy production goes down, step by
step, from the value $\lambda$ to the value $1$ in the long-term regime.
Consider, for instance, the case in which $k>m_{\lambda-1}$. Only in the first
$m_{\lambda-1}$ iteration steps coarse-grained bits (the empty boxes of our
diagram notation) are shifted onto the last bits of the strings
$\y^{j,\lambda-1}$, thereby making them arbitrarily
chose-able for the history, causing a branching into two
alternatives, and increasing the entropy by 1 bit.  In the subsequent
$k-m_{\lambda-1}$ iterations the string $\x^\lambda$ of the
initial condition enters the scale of the $\y^{j,\lambda-1}$-strings, with the
consequence that the last bits of the strings
$\y^{m_{\lambda-1}+1,\lambda-1},\dots,\y^{k,\lambda-1}$ become determined by
the initial condition, meaning no branching and therefore no entropy
increase.

\subsection{Derivation and illustration of the results} 
\label{sec:Derivations}

\subsubsection{1-scale and 2-scale coarse-grainings}
\label{sec:1-scale and 2-scale coarse-grainings}

The decoherence functional for the locally 
coarse-grained histories (\ref{dfunc_local}) was 
calculated in an earlier work of two of the 
authors~\cite{Soklakov2002}. We briefly review 
the corresponding result, which is:  

\begin{equation}              \label{dfunc_local_result}
{\mathcal D}_{B,\,\rho^{\scriptscriptstyle (l,r)}_{\xs}}[h_{\vec{\ys}},
                                    h_{\vec{\zs}}   ]
=2^{-k}
\underbrace{
           \left(\prod_{j=1}^{k}\delta_{\ys^j}^{\zs^j}\right)
}_{{\rm diagonal}}
\cdot
\underbrace{\left(
\delta{}_{\ys^1_{1:\gamma-1}}^{\xs_{2:\gamma}}
\prod_{j=1}^{k-1}
\delta{}_{\ys^{j+1}_{1:\gammas-1}}^{\;\ys^{j}_{2:\gammas}}
\right)}_{{\rm step-by-step\ shift}}
\cdot
\underbrace{ \Bigg{(}
\delta{}_{\ys^k_{1:\gamma-k}}^{\xs_{k+1:\gamma}}
\Bigg{)} }_{k{\rm th\ shift }} 
\;\; +\,O(\frac{l+r-k}{2^{l-2(k^2+k)}})\;, 
\end{equation}  
where $\gamma\equiv |\x|=|\y^j|=|\z^j|=N-(l+r)$.
The expression in the first parentheses is zero for all
off-diagonal elements of the decoherence functional.
In the limit of very large $l$ all off-diagonal elements 
of the decoherence functional vanish, the decoherence condition 
being therefore established. The diagonal elements of the decoherence 
functional can therefore be interpreted as probabilities of the corresponding
histories (see Ref.~\cite{DowkerHalliwell1992} for a discussion of approximate
decoherence). Asymptotically, only $2^{k}$ diagonal elements 
survive. Moreover, the error terms are exponentially small.
We therefore get $2^{k}$ histories with asymptotically
equal probabilities. The number of such histories doubles after 
each iteration step resulting in a loss of information
at the rate of 1 bit per step. This information loss 
is quantified by the entropy increase of the set of histories. 
Since in the limit of large $l$ the set of histories 
$\{h_{\vec{\ys}}\}$ is decoherent, the individual alternative 
histories may be assigned probabilities, which are then given by 
$p[h_{\vec{\ys}}] = {\mathcal D}_{B,\,\rho^{\scriptscriptstyle
    (l,r)}_{\xs}}[h_{\vec{\ys}},h_{\vec{\ys}} ]$. Having found 
the probability distribution we may also define the entropy 
of the set of all possible alternative histories:  
\begin{eqnarray}
H[\{h_{\vec{\ys}} \}] 
 &\equiv&  -\sum_{\vec{\ys}}p[h_{\vec{\ys}}]\log_2p[h_{\vec{\ys}}]\nonumber\\
 &\equiv& -\sum_{\vec{\ys}}{\mathcal D}_{B,\,\rho^{\scriptscriptstyle
    (l,r)}_{\xs}}[h_{\vec{\ys}},h_{\vec{\ys}} ]\log_2
\left({\mathcal D}_{B,\,\rho^{\scriptscriptstyle
    (l,r)}_{\xs}}[h_{\vec{\ys}},h_{\vec{\ys}} ]\right)\;.
\end{eqnarray}
With (\ref{dfunc_local_result}) we obtain:
\begin{equation}               \label{EntropyLocal}
H[\{h_{\vec{\ys}} \}] = k+
O(\frac{(l+r-k)\log_2(l+r-k)}{2^{l-2(k^2+k)}})\,.
\end{equation}
In the limit of large $l$, for any fixed number of iterations, $k$, the
entropy of the coarse-grained quantum baker's map approaches the value of $k$
bits, i.e., 1 bit per iteration. 

What kind of histories arise with significant probabilities? 
This is determined by the expressions within the second and 
third parentheses of the result (\ref{dfunc_local_result}). 
Accordingly only histories that satisfy a
{\em step-by-step shift condition\/}
arise with significant probabilities.
This can be illustrated using the diagram notation introduced above:
\begin{eqnarray}
& \underbrace{\Box\Box\dots\Box}_{l}\;
                     \ \x_1
                  \underline{\x_2\dots\x_{\gamma-2}\x_{\gamma-1} \x_\gamma } \;
   \underbrace{\Box\Box\dots \Box}_{r}\ , \cr
& \ \ \ \ \ \ \boldarrow_{\phantom{|_{|_{|_|}}}} \cr
& \underbrace{\Box\Box\dots\Box}_{l}\;
             \overline{ \ \y^1_1\underline{\y^1_2\dots\y^1_{\gamma-2} \y^1_{\gamma-1}}}
                     \underline{\; y^1_\gamma}\;
   \underbrace{\Box\Box\dots \Box}_{r}\ , \cr
& \ \ \ \ \ \ \boldarrow_{\phantom{|_{|_{|_|}}}} \cr
& \underbrace{\Box\Box\dots\Box}_{l}\;
             \overline{ \ \y^2_1 \underline{\y^2_2\dots\,\y^2_{\gamma-2} y^2_{\gamma-1}}}
                     \underline{\; y^2_\gamma}\;
   \underbrace{\Box\Box\dots \Box}_{r}\ , \cr
& \ \ \ \ \boldarrow \cr
& \ \ \ \ \dots \cr
& \ \ \ \ \ \ \boldarrow_{\phantom{|_{|_{|_|}}}} \cr
& \underbrace{\Box\Box\dots\Box}_{l}\;
             \overline{ \y^k_1 \dots\y^k_{\gamma-k} y^k_{\gamma-k+1}
                       \!\dots}\, y^k_\gamma\,
   \underbrace{\Box\Box\dots \Box}_{r}\ .
\end{eqnarray}
The first line of this diagram represents the initial condition 
$\rho^{{\scriptscriptstyle{(l,r)}}}_{\xs}$. The subsequent lines
correspond to the projectors $P^{(l,r)}_{\ys^1},\dots,P^{(l,r)}_{\ys^k}$
constituting the history $h_{\vec{\ys}} $. 
The step-by-step shift condition is depicted by arrows and lines. 
Underlined substrings are shifted onto those overlined substrings 
which are indicated by arrows.  
In order to fulfil the step-by-step shift condition 
all underlined and overlined substrings that are connected by 
an arrow have to be equal. In this way it becomes clear which 
bits of the symbolic specification of a history are completely 
determined by the initial condition. These bits are in bold face. 
The other bits may be chosen arbitrarily. For instance, in the first iteration 
step the initial condition substring 
$\x_{2:\gamma}\equiv x_2\dots x_{\gamma}$ is 
shifted onto the substring $\y^{1}_{1:\gamma-1}
\equiv y^1_1\dots y^1_{\gamma-1}$.
The first $\gamma-1$ bits of the string $\y^1$ of the 
first event in the history $h_{\vec{\ys}} $  are therefore 
determined by the initial condition. 
Unless $\y^{1}_{1:\gamma-1}=\x_{2:\gamma}$ is satisfied by 
the first event, the whole history will have a vanishing 
probability. On the other hand the last bit $y^{1}_{\gamma}$ 
of the string $\y^{1}$, which denotes the first event of the history, 
remains undetermined, because the unspecified bit of the empty box is 
shifted onto it, which is coarse-grained (i.e.\ summed) over. 
The bit $y^{1}_{\gamma}$ may therefore be chosen arbitrarily, corresponding
to a branching into two possible histories with non-vanishing 
probabilities and therefore an entropy increase of 1 bit.  This 
procedure repeats with each iteration step of the evolution.
For the entire history, therefore, there are only $k$ independent 
bits which can be chosen arbitrarily, given the step-by-step shift 
constraint.

The calculation of the decoherence functional (\ref{checkerboard-dfunc}) 
for the non-locally coarse-grained histories can be traced back to using ´
the above result for the local ones. To do so, we may express 
all the nonlocal projection operators appearing in the 
decoherence functional as sums over suitable local ones: 
\begin{eqnarray}
\rho_{\xs^1,\,\xs^2}^{(l,m_l, m_r, r)}&=& 2^{-(l+m_l+m_r+r)} 
\sum_{\,\bxis\in \{0,1\}^{m_l+m_r}}  {P}_{\xs^1\bxis\xs^2}^{(l,r)}\;,\\
{P}_{\ys^{j,1},\,\ys^{j,2}}^{(l,m_l, m_r, r)}&=&
\sum_{\,\boldetas^j\in \{0,1\}^{m_l+m_r}}
    {P}_{\ys^{j,1}\boldetas^j\ys^{j,2}}^{(l,r)}\;\;,\;\; j=1,2,\dots,k\;,\\
{P}_{\zs^{j,1},\,\zs^{j,2}}^{(l,m_l, m_r, r)}&=&
\sum_{\,\bzetas^j\in \{0,1\}^{m_l+m_r}}
    {P}_{\zs^{j,1}\bzetas^j\zs^{j,2}}^{(l,r)}\;\;,\;\;\; j=1,2,\dots,k\;.
\end{eqnarray}
By inserting these expressions into the decoherence functional 
(\ref{checkerboard-dfunc}) we arrive at:
\begin{eqnarray}
&&\hspace*{-1.3cm}
{\mathcal D}_{B,\,\rho_0}[h_{\vec{\ys}^1,\,\vec{\ys}^2},
  h_{\vec{\zs}^1,\,\vec{\zs}^2} ]=\nonumber\\&=&
\sum_{\,\boldetas^1\in \{0,1\}^{m_l+m_r}}\dots
\sum_{\,\boldetas^k\in \{0,1\}^{m_l+m_r}}
\sum_{\,\bxis\in \{0,1\}^{m_l+m_r}}
\sum_{\,\bzetas^1\in \{0,1\}^{m_l+m_r}}\dots
\sum_{\,\bzetas^k\in \{0,1\}^{m_l+m_r}}
\nonumber\\&&
2^{-(m_l+m_r)}\Tr\Big[ \phantom{\Big]}
{P}_{\ys^{k,1}\boldetas^k\ys^{k,2}}^{(l,r)}B
                    {P}_{\ys^{k-1,1}\boldetas^{k-1}\ys^{k-1,2}}^{(l,r)} 
                    B\cdots B
                    {P}_{\ys^{1,1}\boldetas^1\ys^{1,2}}^{(l,r)}\times \nonumber \\
                     &&\times   \phantom{\Big[} B
                    \big(\frac{1}{2^{l+r}}{P}_{\xs^1\xibolds\xs^2}^{(l,r)} \big)
                    B^\dag {P}_{\zs^{1,1}\bzetas^1\zs^{1,2}}^{(l,r)}
                    B^\dag\cdots
                     {P}_{\zs^{k-1,1}\bzetas^{k-1}\zs^{k-1,2}}^{(l,r)}B^\dag
                     {P}_{\zs^{k,1}\bzetas^{k}\zs^{k,2}}^{(l,r)}\Big]\nonumber\\&&
\end{eqnarray}
Each term of the sum over all possible strings $\bxi$, $\{\boldeta^j\}$
and $\{\bzeta^j\}$ is, apart from the factor $2^{-(m_l+m_r)}$, a
decoherence functional with respect to histories composed of 
local projectors. Each such term, therefore, results in an expression 
of the form (\ref{dfunc_local_result}), and we obtain: 
\begin{eqnarray}
&&\hspace*{-1.3cm}
{\mathcal D}_{B,\,\rho_0}[h_{\vec{\ys}^1,\,\vec{\ys}^2},
  h_{\vec{\zs}^1,\,\vec{\zs}^2} ]=\nonumber\\&=&
\sum_{\,\boldetas^1\in \{0,1\}^{m_l+m_r}}\dots
\sum_{\,\boldetas^k\in \{0,1\}^{m_l+m_r}}
\sum_{\,\bxis\in \{0,1\}^{m_l+m_r}}
\sum_{\,\bzetas^1\in \{0,1\}^{m_l+m_r}}\dots
\sum_{\,\bzetas^k\in \{0,1\}^{m_l+m_r}}2^{-(m_l+m_r)}\times\nonumber\\&&
\times\Bigg\{  \phantom{\Bigg\} }  2^{-k}  
\underbrace{
\left(\prod_{i=1}^{k}\delta_{\ys^{i,1}\boldetas^i\ys^{i,2}}
                           ^{\zs^{i,1}\bzetas^i\zs^{i,2}}\right)
}_{{\rm diagonal}}\cdot
\underbrace{\left(
\delta{}_{\ys^{1,1}\boldetas^1\ys^{1,2}_{1:s_2-1}}
^{\xs^1_{2:s_1}\bxis\,\xs^2}
\prod_{j=1}^{k-1}
\delta{}_{\ys^{j+1,1}\boldetas^{j+1}\ys^{j+1,2}_{1:(s_2-1)}}
^{\ys^{j,1}_{2:s_1}\boldetas^{j}\ys^{j,2}}\right)}_{{\rm step-by-step\
  shift}}\times
\nonumber\\&&\times \phantom{\Bigg\{ }
\underbrace{ \Bigg{(}
\delta{}_{(\ys^{k,1}\boldetas^k\ys^{k,2})_{1:\gamma-k}}
^{(\xs^1\bxis\,\xs^2)_{k+1:\gamma}}
\Bigg{)} }_{k{\rm th\ shift }} 
\;\; +\,{\cal O}(\frac{l+r-k}{2^{l-2(k^2+k)}})\;
\Bigg\}\;\nonumber\\
&=&
\sum_{\,\boldetas^1\in \{0,1\}^{m_l+m_r}}\dots
\sum_{\,\boldetas^k\in \{0,1\}^{m_l+m_r}}
\sum_{\,\bxis\in \{0,1\}^{m_l+m_r}}\Bigg\{  \phantom{\Bigg\} }
2^{-(m_l+m_r)}\cdot 2^{-k}\cdot\underbrace{
\left(\prod_{i=1}^{k}\delta_{\ys^{i,1}\boldetas^i\ys^{i,2}}
                           ^{\zs^{i,1}\boldetas^i\zs^{i,2}}\right)
}_{{\rm diagonal}}\times\nonumber\\ &&\times
\underbrace{\left(
\delta{}_{\ys^{1,1}}^{\xs^1_{2:s_1}\xi_1}
\delta{}_{\boldetas^1}^{\bxis_{2:(m_l+m_r)}\xs^2_1}
\delta{}_{\ys^{1,2}_{1:s_2-1}}^{\xs^2_{2:s_2}}\cdot
\prod_{j=1}^{k-1}
\delta{}_{\ys^{j+1,1}}^{\ys^{j,1}_{2:s_1}\eta^j_1}
\delta{}_{\boldetas^{j+1}}^{\boldetas^{j}_{2:(m_l+m_r)}\ys^{j,2}_1}
\delta{}_{\ys^{j+1,2}_{1:s_2-1}}^{\ys^{j,2}_{2:s_2}}
\right)}_{{\rm step-by-step\ shift}}\times \nonumber\\&&\times
\underbrace{ \Bigg{(}
\delta{}_{(\ys^{k,1}\boldetas^k\ys^{k,2})_{1:\gamma-k}}
^{(\xs^1\bxis\,\xs^2)_{k+1:\gamma}}
\Bigg{)} }_{k{\rm th\ shift }} \phantom{\Bigg\{ } \Bigg\}
\;\; +\,\,\Big( 2^{m_l+m_r}\Big)^{2k}\cdot\, 
{\cal O}(\frac{l+r-k}{2^{l-2(k^2+k)}})\;
\nonumber\\
\end{eqnarray}
Here $\gamma$ denotes the length of the strings 
$\y^{j,1}\boldeta^j\y^{j,2}$ and 
$\x^1\bxi\,\x^2$, respectively, 
i.e.\  $\gamma=|\x^1\bxi\,\x^2|=|\y^{j,1}\boldeta^j\y^{j,2}|=
s_1+(m_l+m_r)+s_2$. 

First of all the sum over all possible $\bzeta^j\in
\{0,1\}^{m_l+m_r}$, $j=1,\dots, k$, collapses due to the term 
$\prod_{i=1}^{k}\delta_{\ys^{i,1}\boldetas^i\ys^{i,2}}^
{\zs^{i,1}\bzetas^i\zs^{i,2}}\,$, apart from contributing a 
factor $2^{k(m_l+m_r)}$ to the error term. Secondly we note 
that the step-by-step shift condition causes the whole sum 
$\sum_{\boldetas^1}\sum_{\boldetas^2}\dots\sum_{\boldetas^k}$ 
to collapse, apart from contributing a further factor 
$2^{k(m_l+m_r)}$ to the bound on the error term, 
which is furthermore enlarged  by a factor $2^{(m_l+m_r)}$ 
stemming from the sum $\sum_{\bxis}$. Let us comprehend the 
collapse of the sums $\sum_{\boldetas^j}$. For a given fixed string 
$\bxi$ out of the sum  $\sum_{\bxis}$ all $\boldeta^1, \boldeta^2, 
\dots, \boldeta^k$ are through the $\delta$'s determined by the 
string $\bxi$ and the given fixed string $\x^2$ of the initial 
condition. The first shift leads to a determination of $\boldeta^1$: 
according to $\delta{}_{\boldetas^1}^{\bxis_{2:(m_l+m_r)}\xs^2_1}$ 
the sum over all possible $\boldeta^1\in\{0,1\}^{m_l+m_r}$ collapses 
and only the string $\boldeta^1=\hspace{-3.3mm}^!\;\;
\bxi_{2:(m_l+m_r)}\x^2_1$ survives. The second shift determines 
$\boldeta^2$, since according to $\delta{}_{\boldetas^{2}}^
{\boldetas^{1}_{2:(m_l+m_r)}\ys^{1,2}_1}$ the sum over all possible
$\boldeta^2\in\{0,1\}^{m_l+m_r}$ collapses and only the string 
$\boldeta^2= \;\boldeta^1_{2:(m_l+m_r)}\y^{1,2}_1
\equiv\bxi_{3:(m_l+m_r)}\x^2_1\x^2_2$ does lead to a non-vanishing 
contribution to the decoherence functional. It is easy to see that 
due to the step-by-step shift condition all the sums
$\sum_{\boldetas^j}$, $j=1,\dots, k$, collapse and only the strings 
\begin{eqnarray}\label{determinationetas}
\boldeta^j=\bxi_{(j+1):(m_l+m_r)}\x^2_{1:j}
\end{eqnarray}
out of these sums survive leading together to a non-vanishing 
contribution to the decoherence functional. In fact the step-by-step 
shift condition can also be expressed in the following way:
\begin{equation}
\prod_{j=1}^{k} \delta{}_{(\ys^{j,1}\boldetas^j\ys^{j,2})_{1:\gamma-j}}
^{(\xs^1\bxis\,\xs^2)_{j+1:\gamma}}\;\:,
\end{equation}
meaning that only such strings $\boldeta^j$ out of the
corresponding sums $\sum_{\boldetas^j}$, $j=1,\dots, k$, 
lead to a non-vanishing contribution to the decoherence functional 
which are determined by $\bxi$ and $\x^2$ according to (\ref{determinationetas}).
Next we note that as a consequence of the step-by-step shift condition  
also the sum over all possible $\bxi\in\{0,1\}^{m_l+m_r}$ collapses. 
It collapses only {\em partially} in case $k< m_l+m_r$ and it
collapses {\em completely} in case $k \ge m_l+m_r$. Let us first consider 
the case $k< m_l+m_r$. After the first shift the first bit of $\bxi$ 
is determined by the last bit of the string $\y^{1,1}$ of the given 
history, i.e.\ $\xi_1=\hspace{-3.4mm}^!\;\;y^{1,1}_{s_1}$, 
according to the term  $\delta{}_{\ys^{1,1}}^{\xs^1_{2:s_1}\xi_1}$. 
The second shift leads to
$\delta{}_{\ys^{2,1}}^{\ys^{1,1}_{2:s_1}\eta^1_1}$, 
so that $\eta_1^1= \;y^{2,1}_{s_1}$. 
But we have $\eta_1^1=  \;\xi_2$ due to the 
first shift, so we arrive at a determination of  $\xi_2$, 
namely $\xi_2= \;y^{2,1}_{s_1}$. In this 
way the sum over all possible $\bxi\in\{0,1\}^{m_l+m_r}$ collapses 
to a sum over all possible
$\bxi_{(k+1):(m_l+m_r)}\in\{0,1\}^{m_l+m_r-k}$, 
\begin{equation}
\sum_{\,\bxis\in \{0,1\}^{m_l+m_r}}\longrightarrow 
\sum_{\,\bxis_{(k+1):(m_l+m_r)}\in\{0,1\}^{m_l+m_r-k}}\;,
\end{equation}
since the first $k$ bits $\xi_1, \dots, \xi_k$ out of the sum $\sum_{\bxis}$ 
have to fulfil the step-by-step shift condition and are therefore 
determined by $\xi_j= \;y^{j,1}_{s_1}$. 
That the first $k$ bits of the string $\bxi$ out of the sum
$\sum_{\bxis}$ are determined by the given history 
$h_{\vec{\ys}^1,\,\vec{\ys}^2}$ and therefore the sum over 
the first  $k$ bits of $\bxi= \xi_1\xi_2\dots\xi_{m_l+m_r}$ 
collapses can also be seen by looking at the $k$-th shift factor 
which in fact appears as a redundant factor in the result: according 
to $ \delta{}_{(\ys^{k,1}\boldetas^k\ys^{k,2})_{1:\gamma-k}}
^{(\xs^1\bxis\,\xs^2)_{k+1:\gamma}}$ only such strings $\bxi$ 
out of the sum $\sum_{\bxis}$ lead to a non-vanishing contribution to
the decoherence functional for which $\bxi_{1:k}= \;
\y^{k,1}_{(s_1-k+1):s_1}$ holds. The remaining $m_l+m_r-k$ bits of 
$\bxi= \xi_1\xi_2\dots\xi_{m_l+m_r}$ remain undetermined and are 
still summed over. There are $2^{m_l+m_r-k}$ possible different 
substrings $\bxi_{k+1:m_l+m_r}\in\{0,1\}^{m_l+m_r-k}$ in this 
remaining sum leading to a non-vanishing contribution to the
decoherence functional. Since the contributions of all these strings 
are equal, as can be seen by looking at the result, we may replace
the remaining sum over all possible $\bxi_{k+1:m_l+m_r}$ by the factor 
$2^{m_l+m_r-k}$. Furthermore all the $\delta$-terms containing bits of the
unspecified strings $\bxi$ and $\boldeta^j$, $j=1,\dots,k$, which are 
summed over, may now be replaced by $1$ after having been exploited 
for the determination of that strings $\bxi$ and $\boldeta^j$ out 
of the sums $\sum_{\bxis}$ and $\sum_{\boldetas^1}\sum_{\boldetas^2}
\dots\sum_{\boldetas^k}$ which lead to a non-vanishing contribution 
to the value of the decoherence functional. In case $k<m_l+m_r$ we 
therefore arrive at the following result:

\begin{itemize}
\item Decoherence functional in case  $k<m_l+m_r$:
\begin{eqnarray}\label{result_k<(m_l+m_r)}
{\mathcal D}_{B,\,\rho_0}[h_{\vec{\ys}^1,\,\vec{\ys}^2},
  h_{\vec{\zs}^1,\,\vec{\zs}^2} ]&=&
\underbrace{2^{m_l+m_r-k}\cdot 2^{-(m_l+m_r)}\cdot 2^{-k}}_{=\,2^{-2k}}
\cdot\underbrace{\left(\prod_{i=1}^{k}\delta_{\ys^{i,1}}^{\zs^{i,1}}
\delta_{\ys^{i,2}}^{\zs^{i,2}}\right)}_{{\rm diagonal}}\times\nonumber\\ 
&&\times
\underbrace{\left(
\delta{}_{\ys^{1,1}_{1:s_1-1}}^{\xs^1_{2:s_1}}
\delta{}_{\ys^{1,2}_{1:s_2-1}}^{\xs^2_{2:s_2}}\cdot
\prod_{j=1}^{k-1}
\delta{}_{\ys^{j+1,1}_{1:s_1-1}}^{\ys^{j,1}_{2:s_1}}
\delta{}_{\ys^{j+1,2}_{1:s_2-1}}^{\ys^{j,2}_{2:s_2}}
\right)}_{{\rm step-by-step\ shift}}\times \\&&\times
\underbrace{ \Bigg{(}
\delta{}_{\ys^{k,1}_{1:s_1-k}}^{\xs^1_{k+1:s_1}}
\delta{}_{\ys^{k,2}_{1:s_2-k}}^{\xs^2_{k+1:s_2}}
\Bigg{)} }_{k{\rm th\ shift }}
\;\; +\,\, {\cal
  O}(\frac{l+r-k}{2^{l-2(k^2+(1+m_l+m_r)k)}})\;.
\nonumber
\end{eqnarray}
\end{itemize}

\noindent
Let us now consider the case $k\ge m_l+m_r$. As already mentioned in
this case the whole sum $\sum_{\bxis}$ collapses to a single string 
$\bxi\in\{0,1\}^{m_l+m_r}$ satisfying the step-by-step shift
condition. This can be seen, again, by looking at the $k$-th shift 
condition  given by the factor $\delta{}_{(\ys^{k,1}\boldetas^k
\ys^{k,2})_{1:\gamma-k}}^{(\xs^1\bxis\,\xs^2)_{k+1:\gamma}}$; 
according to it each string $\bxi$ out of the sum $\sum_{\bxis}$ 
is shifted onto $(m_l+m_r)$ bits of the string $\y^{k,1}$, but since 
$\y^{k,1}$ is a {\em fixed} string specifying the last event of the 
{\em given} history, only  the string 
$\bxi= \;\y^{k,1}_{(s_1-k+1):(s_1-k+m_l+m_r)}$ 
out of the sum  $\sum_{\bxis}$ survives. Of course we presupposed,
or had to require, that $s_1\ge k\ge m_l+m_r$.
In case $k\ge m_l+m_r$ we therefore get:

\begin{itemize}
\item Decoherence functional in case  $k\ge m_l+m_r$:
\begin{eqnarray} \label{result_k>(m_l+m_r)}
{\mathcal D}_{B,\,\rho_0}[h_{\vec{\ys}^1,\,\vec{\ys}^2},
  h_{\vec{\zs}^1,\,\vec{\zs}^2} ]&=&
2^{-(m_l+m_r)} \cdot 2^{-k}
\cdot\underbrace{\left(\prod_{i=1}^{k}\delta_{\ys^{i,1}}^{\zs^{i,1}}
\delta_{\ys^{i,2}}^{\zs^{i,2}}\right)}_{{\rm
  diagonal}}\times\nonumber\\&& 
\times
\underbrace{\left(
\delta{}_{\ys^{1,1}_{1:s_1-1}}^{\xs^1_{2:s_1}}
\delta{}_{\ys^{1,2}_{1:s_2-1}}^{\xs^2_{2:s_2}}\cdot
\prod_{j=1}^{k-1}
\delta{}_{\ys^{j+1,1}_{1:s_1-1}}^{\ys^{j,1}_{2:s_1}}
\delta{}_{\ys^{j+1,2}_{1:s_2-1}}^{\ys^{j,2}_{2:s_2}}
\right)}_{{\rm step-by-step\ shift}}\times \\&&\times
\underbrace{ \Bigg{(}
\delta{}_{\ys^{k,1}_{1:s_1-k}}^{\xs^1_{k+1:s_1}}
\delta{}_{\ys^{k,1}_{s_1-k+(m_l+m_r)+1\,:\,s_1}}^{\xs^2_{1:\,k-(m_l+m_r)}}
\delta{}_{\ys^{k,2}_{1:s_2-k}}^{\xs^2_{k+1:s_2}}
\Bigg{)} }_{k{\rm th\ shift }}\nonumber\\
&&\;\; +\,\, {\cal
  O}(\frac{l+r-k}{2^{l-2(k^2+(1+m_l+m_r)k)}})\;.
 \nonumber
\end{eqnarray}
\end{itemize}

Let us now discuss the  results (\ref{result_k<(m_l+m_r)}) 
and (\ref{result_k>(m_l+m_r)}) for the decoherence functional 
(\ref{checkerboard-dfunc}). First of all we get approximate 
decoherence: for very large $l$ the decoherence functional is 
approximately diagonal. In the asymptotic limit 
$l \rightarrow \infty $ our set of histories 
$\{ h_{\vec{\ys}^1,\,\vec{\ys}^2}\}$ becomes decoherent. 
The diagonal elements of the functional, 
$\mathcal{D}_{B,\,\rho_0}[h_{\vec{\ys}^1,\,\vec{\ys}^2}, 
h_{\vec{\ys}^1,\,\vec{\ys}^2} ]$, may therefore be interpreted 
as probabilities of the corresponding histories, i.e.\ 
$p(h_{\vec{\ys}^1,\,\vec{\ys}^2},)=
\mathcal{D}_{B,\,\rho_0}[h_{\vec{\ys}^1,\,\vec{\ys}^2}, 
h_{\vec{\ys}^1,\,\vec{\ys}^2} ]$. 
Again there is no single dominant history. 
Several different histories arise with significant probabilities$\,$. 
In case $k<m_l+m_r$ we get $2^{2k}$ different histories with
asymptotically equal probabilities (given by $2^{-2k}$). The number 
of histories with asymptotically nonzero probabilities becomes 
four times larger after each iteration step of the quantum 
baker's map resulting in a {\em loss of information of $2$ bits per step}. 
The entropy increase is therefore {\em $2$  bits per iteration step}, 
which can also be seen by calculating the entropy of the 
approximately decoherent set of histories $\{ h_{\vec{\ys}^1,\,\vec{\ys}^2}\}$:

\begin{itemize}
\item Entropy after $k$ iteration steps in case  $k\le m_l+m_r$:
\begin{eqnarray}
H[\{ h_{\vec{\ys}^1,\,\vec{\ys}^2}\}] 
& = & -\sum_{\vec{\ys}^1,\,\vec{\ys}^2 }p[ h_{\vec{\ys}^1,\,\vec{\ys}^2}]
\log_2p[h_{\vec{\ys}^1,\,\vec{\ys}^2}]\nonumber\\
 &=& 
2k\;+\;{\cal
  O}(\frac{(l+r-k)\log_2(l+r-k)}{2^{l-2(k^2+(1+m_l+m_r)k)}})\;.
\end{eqnarray}
\end{itemize}

\noindent
Again, only such histories are allowed to arise with significant 
probabilities that satisfy the shift condition: the projectors 
of the histories have to be related to the initial state via a shift. 
Let us illustrate this issue once again by means of our diagram
notation:

\begin{eqnarray}
& \underbrace{\Box\Box\dots\Box}_{l}\;
                     \ \x_1^1
                  \underline{\x^1_2\dots\x^1_{s_1-2}\x^1_{s_1-1}
                     \x^1_{s_1}}\;\;\underbrace{\Box\Box\dots\Box}_{m_l+m_r} \;\ \x^2_1
                  \underline{\x^2_2\dots\x^2_{s_2-2}\x^2_{s_2-1}\x^2_{s_2}} \;
   \underbrace{\Box\Box\dots \Box}_{r}\  \cr
&\hspace{-0.5cm} \ \ \ \ \ \ \boldarrow_{\phantom{|_{|_{|_|}}}} 
\hspace{4.5cm}\ \ \ \ \ \ \boldarrow_{\phantom{|_{|_{|_|}}}}
\cr
& \underbrace{\Box\Box\dots\Box}_{l}\;
             \overline{ \
             \y^{1,1}_1\underline{\y^{1,1}_2\dots\y^{1,1}_{s_1-2}
             \y^{1,1}_{s_1-1}}}  \underline{\; y^{1,1}_{s_1}}\;\;
\underbrace{\Box\Box\dots\Box}_{m_l+m_r} \;\  \overline{ \
\y^{1,2}_1\underline{\y^{1,2}_2\dots\y^{1,2}_{s_2-2}
 \y^{1,2}_{s_2-1}}}  \underline{\; y^{1,2}_{s_2}}\;\;
   \underbrace{\Box\Box\dots \Box}_{r}\  \cr
&\hspace{-0.5cm} \ \ \ \ \ \ \boldarrow_{\phantom{|_{|_{|_|}}}} 
\hspace{5.5cm}\ \ \ \ \ \ \boldarrow_{\phantom{|_{|_{|_|}}}}\cr
& \underbrace{\Box\Box\dots\Box}_{l}\;
             \overline{ \
             \y^{2,1}_1\underline{\y^{2,1}_2\dots\y^{2,1}_{s_1-2}
             y^{2,1}_{s_1-1}}}  \underline{\; y^{2,1}_{s_1}}\;\;
\underbrace{\Box\Box\dots\Box}_{m_l+m_r} \;\  \overline{ \
\y^{2,2}_1\underline{\y^{2,2}_2\dots\y^{2,2}_{s_2-2}
 y^{2,2}_{s_2-1}}}  \underline{\; y^{2,2}_{s_2}}\;\;
   \underbrace{\Box\Box\dots \Box}_{r}\  \cr
&\hspace{-0.5cm} \ \ \ \ \ \ \boldarrow_{\phantom{|_{|_{|_|}}}} 
\hspace{5.5cm}\ \ \ \ \ \ \boldarrow_{\phantom{|_{|_{|_|}}}}\cr
& \hspace{-0.5cm}\ \ \ \ \dots\hspace{6cm}\ \ \ \ \ \dots  \cr
&\hspace{-0.5cm} \ \ \ \ \ \ \boldarrow_{\phantom{|_{|_{|_|}}}} 
\hspace{5.5cm}\ \ \ \ \ \ \boldarrow_{\phantom{|_{|_{|_|}}}}\cr
& \underbrace{\Box\Box\dots\Box}_{l}\;
 \overline{ \y^{k,1}_1 \dots\y^{k,1}_{s_1-k} y^{k,1}_{s_1-k+1}
                       \!\dots}\, y^{k,1}_{s_1}\;\;
\underbrace{\Box\Box\dots\Box}_{m_l+m_r} \;\  
 \overline{ \y^{k,2}_1 \dots\y^{k,2}_{s_2-k} y^{k,2}_{s_2-k+1}
                       \!\dots}\, y^{k,2}_{s_2}\;\;
   \underbrace{\Box\Box\dots \Box}_{r}\  \cr \nonumber
\end{eqnarray}
\begin{equation}
\end{equation}

This diagram illustrates symbolically the content of the result
(\ref{result_k<(m_l+m_r)}). Again, the first line of this diagram represents
the initial condition $\rho_0 = \rho_{\xs^1,\,\xs^2}^{{\scriptstyle{(l,m_l,
      m_r, r)}}}$.  The subsequent lines correspond to the projectors
${P}_{\ys^{1,1},\,\ys^{1,2}}^{(l,m_l, m_r, r)},\dots,
{P}_{\ys^{k,1},\,\ys^{k,2}}^{(l,m_l, m_r, r)}$ representing the subsequent
propositions of the history $ h_{\vec{\ys}^1,\,\vec{\ys}^2}$. The
coarse-grained islands in the middle of each line, with $m_l+m_r$ empty boxes
each, subsequently represent the sums $\sum_{\bxis}$, $\sum_{\boldetas^1},
\sum_{\boldetas^2},\dots\,\sum_{\boldetas^k}$ in our calculation.  Again, the
step-by-step shift condition is depicted by arrows and lines.  Underlined
substrings are shifted onto those overlined substrings which are indicated by
arrows.  In order to fulfil the step-by-step shift condition all underlined
and overlined substrings that are connected by an arrow must be equal. In this
way we immediately see which bits of the symbolic specification of a history
are completely determined by the initial condition. In the diagram these bits
are indicated by using bold face. The remaining bits, which are not in bold
face, may be chosen arbitrarily. For instance, in the first iteration step the
initial condition substrings $\;\x^1_{2:s_1}\equiv x^1_2\dots x^1_{s_1-2}
x^1_{s_1-1} x^1_{s_1}\;$ and $\;\x^2_{2:s_2}\equiv x^2_2\dots x^2_{s_2-2}
x^2_{s_2-1} x^2_{s_2}\;$ are shifted onto the substrings
$\;\y^{1,1}_{1:(s_1-1)}\equiv y^{1,1}_1 y^{1,1}_2\dots
y^{1,1}_{s_1-2}y^{1,1}_{s_1-1}\;$ and $\;\y^{1,2}_{1:(s_2-1)}\equiv y^{2,2}_1
y^{2,2}_2\dots y^{2,2}_{s_2-2}y^{2,2}_{s_2-1}\;$, respectively. The first
$(s_1-1)$ bits of the string $\y^{1,1}$ and the first $(s_2-1)$ bits of the
string $\y^{1,2}$ of the first event in the history are therefore determined
by the initial condition. Unless $\y^{1,1}_{1:(s_1-1)}=\x^1_{2:s_1}$ and
$\y^{1,2}_{1:(s_2-1)}=\x^2_{2:s_2}$ is satisfied by the first event the whole
history will have  vanishing probability. On the other hand the last bits
$y^{1,1}_{s_1}$ and $y^{1,2}_{s_2}$ of the strings $\y^{1,1}$ and $\y^{1,2}$,
which denote the first event of the history, remain undetermined, because the
unspecified bits of the empty boxes are shifted onto them, which are
coarse-grained (i.e.\ summed) over. The bits $y^{1,1}_{s_1}$ and
$y^{1,2}_{s_2}$ may therefore be chosen arbitrarily resulting in a branching into
four possible histories with non-vanishing probabilities.  This procedure
repeats with each iteration step of the evolution. The second step leads to a
determination of the first $(s_1-1)$ bits of the string $\y^{2,1}$ and the
first $(s_2-1)$ bits of the string $\y^{2,2}$ symbolising the second event of
the history, whereas, again, the last bits of these strings remain unspecified
and may be chosen arbitrarily implicating a branching into further four
alternatives with non-vanishing probabilities. And so on. It becomes clear
from the above picture which histories arise with significant probabilities
during the evolution and why the number of alternative equiprobable histories
is quadruplicated after each iteration step. After $k$ iteration steps---we
still assume $k<m_l+m_r$---there are therefore $2k$ independent bits which can
be chosen arbitrarily, given the step-by-step shift constraint.  This
implicates $2^{2k}$ alternative, equiprobable histories that may arise with
significant probability after $k$ iteration steps.

Our result for $k>m_l+m_r$, Eq.\ (\ref{result_k>(m_l+m_r)}),
 may be interpreted in the following way. 
As long as the number of iterations $k$ is smaller than $m=m_l+m_r$ 
the number of histories with asymptotically non-vanishing
probabilities becomes four times larger after each iteration step of 
the quantum baker's map resulting in an entropy increase 
of $2$ bits per iteration 
step. As soon as the number of iterations becomes greater than 
$m=m_l+m_r$, the entropy increase becomes $1$ bit per iteration 
step. This is what is expressed by the result $2^{-(m_l+m_r)}\cdot 
2^{-k}= 2^{-2(m_l+m_r)}\cdot 2^{-(k-(m_l+m_r))}$ for the probability 
of the histories which are allowed to occur. The first  
$m_l+m_r$ iteration steps lead to an entropy increase of $2$ bits 
per step involving $2^{2(m_l+m_r)}$ asymptotically equiprobable 
histories. The remaining $k-(m_l+m_r)$ iteration steps produce 
an entropy increase of $1$ bit per step only, with the number of 
histories with significant probabilities being doubled at each step,
implicating  a branching factor $2^{k-(m_l+m_r)}$. The entire number 
of histories arising with significant probabilities after 
$k$ iteration steps therefore becomes $2^{2(m_l+m_r)}\cdot 2^{k-(m_l+m_r)}
=2^{(m_l+m_r)}\cdot 2^k$, the histories being asymptotically
 equiprobable. Again the issue becomes clearer when using our diagram 
picture. The size of the middle coarse-grained islands 
is now only   $m_l+m_r<k$. So only in the first $m_l+m_r$ iteration 
steps coarse-grained  bits are shifted onto the last bits 
of the strings $\y^{j,1}$, making them by this means unspecified, i.e.\  
arbitrarily chose-able for the history. In the subsequent, 
remaining $k-(m_l+m_r)$ iteration steps the string $\x^2$ of the 
initial condition enters the scale of the $\y^{j,1}$-strings,  
with the consequence that the last bits of the strings  
$\y^{m_l+m_r+1,1},\dots,\y^{k,1}$ 
become determined by the initial condition. 
At the end, after the $k$-th iteration step, only  $m_l+m_r$ bits 
of the string $\y^{k,1}$ may be chosen arbitrarily, the first 
$s_1-k$ bits and the last $k-(m_l+m_r)$ bits of it being determined 
by the initial condition. On the other hand only the first $s_2-k$ 
bits of the string $\y^{k,2}$ become determined by the initial
condition, whereas all the last $k$ bits of it remain 
arbitrarily chose-able for the history, provided that $k<r$. 
This explains the result  $2^{(m_l+m_r)}\cdot 2^k$ for the 
number of alternative histories satisfying the shift constraint. 
For the entropy of the approximately 
decoherent set of histories $\{ h_{\vec{\ys}^1,\,\vec{\ys}^2}\}$
we get the result:
\begin{itemize}
\item Entropy after $k$ iteration steps in case  $k\ge m_l+m_r$:
\begin{eqnarray}
H[\{ h_{\vec{\ys}^1,\,\vec{\ys}^2}\}] 
 &=& -\sum_{\vec{\ys}^1,\,\vec{\ys}^2 }p[ h_{\vec{\ys}^1,\,\vec{\ys}^2}]
\log_2p[h_{\vec{\ys}^1,\,\vec{\ys}^2}] \nonumber \\
&=&k+\;(m_l+m_r)\;+\;{\cal
  O}(\frac{(l+r-k)\log_2(l+r-k)}{2^{l-2(k^2+(1+m_l+m_r)k)}})\;.
\end{eqnarray}
\end{itemize}

\subsubsection{Hierarchical (multi-scale) coarse-grainings} 
\label{sec:Derivations_Hierarchical_coarse-grainings}

We will see in the following that by introducing more and more 
scales that are coarse-grained over in the symbolic representation of 
the dynamics the short-term behaviour of the coarse-grained 
evolution of the quantum baker's map will exhibit a growing 
entropy increase per iteration step, i.e., growing unpredictability. 

So let us now look at the generalised type of 
histories~(\ref{set:hierarchically-coarse-grained-histories}). 
The evaluation of the corresponding decoherence 
functional~(\ref{hierarchical-dfunc})  is done in a similar
way as for the case $\lambda=2$. We first state the result for 
the short-term regime which we now define to be given by 
$k<\mbox{min}\{m_1,m_2, \dots, m_{\lambda-1}\}$:  
\begin{itemize}
\item Decoherence functional in the case  
$k<\mbox{min}\{m_1,m_2, \dots, m_{\lambda-1}\}$:
\begin{eqnarray}\label{result_[k<min_m_j]}
{\mathcal D}_{B,\,\rho_0}
[h_{\vec{\ys}^1,\,\vec{\ys}^2,\dots,\vec{\ys}^{\lambda}}\,,\,
h_{\vec{\zs}^1,\,\vec{\zs}^2,\dots,\vec{\zs}^{\lambda}}]
&=&
2^{-\lambda k}
\cdot\underbrace{\left(\,\prod_{j=1}^{k}
\prod_{i=1}^{\lambda}
\delta_{\ys^{j,i}}^{\zs^{j,i}}
\right)}_{{\rm diagonal}}\cdot
\underbrace{\left(\prod_{i=1}^{\lambda}
\delta{}_{\ys^{1,i}_{1:s_i-1}}^{\xs^i_{2:s_i}}
\right)}_{{\rm first \; shift}}\times
\nonumber\\
&& \times
\underbrace{\left(\,\prod_{j=1}^{k-1}\prod_{i=1}^{\lambda}
\delta{}_{\ys^{j+1,i}_{1:s_i-1}}^{\ys^{j,i}_{2:s_i}}
\right)}_{{\rm step-by-step\ shift}}\cdot
\underbrace{\left(\prod_{i=1}^{\lambda}
\delta{}_{\ys^{k,i}_{1:s_i-k}}^{\xs^i_{k+1:s_i}}
\right)}_{k{\rm -th\ shift }}\\&&
\;\; +\,\, {\cal
  O}\Big(\frac{l+r-k}{2^{l-2(k^2+(1+m_1+m_2+\dots+m_{\lambda-1})k)}}\Big)\;.
\nonumber
\end{eqnarray}
\end{itemize}

In the limit of large $l$ the off-diagonal elements of the 
decoherence functional vanish and the set of histories becomes 
decoherent. The diagonal elements of the functional may therefore 
be interpreted as probabilities. The coarse-grained evolution is 
again governed by shift constraints. Only such histories are allowed 
to arise with significant probabilities that satisfy the shift
condition, which has been illustrated in detail for the case
$\lambda=2$ above. Here we are mainly interested 
in the rate of the entropy increase. The result
(\ref{result_[k<min_m_j]}) shows that in the short-term regime, 
i.e.\ as long as $k<\mbox{min}\{m_1,m_2, \dots, m_{\lambda-1}\}$, 
the coarse-grained evolution exhibits 
an entropy increase of $\lambda$ bits per iteration step, 
provided that $l$ is very large (classical limit).  
This is quantitatively expressed by the entropy of the  
approximately decoherent set of histories:  

\begin{itemize}
\item Entropy after $k$ iteration steps in case  
$k<\mbox{min}\{m_1,m_2, \dots, m_{\lambda-1}\}$:
\begin{eqnarray}
H[\{ 
h_{\vec{\ys}^1,\,\vec{\ys}^2,\dots,\vec{\ys}^{\lambda}}\}] 
& = & -\sum_{\vec{\ys}^1,\,\vec{\ys}^2,\dots,\vec{\ys}^{\lambda}}
p[ h_{\vec{\ys}^1,\,\vec{\ys}^2,\dots,\vec{\ys}^{\lambda}}]
\log_2
p[ h_{\vec{\ys}^1,\,\vec{\ys}^2,\dots,\vec{\ys}^{\lambda}}]
\nonumber\\ 
&=&
\lambda\cdot k\;+\;{\cal
  O}(\frac{(l+r-k)\log_2(l+r-k)}{2^{l-2(k^2+(1+m_1+m_2+\dots+m_{\lambda-1})k)}})\;,
\end{eqnarray}
\end{itemize}
where we used $p[ h_{\vec{\ys}^1,\,\vec{\ys}^2,\dots,\vec{\ys}^{\lambda}}]
={\mathcal D}_{B,\,\rho_0}
[h_{\vec{\ys}^1,\,\vec{\ys}^2,\dots,\vec{\ys}^{\lambda}}\,,\,
h_{\vec{\ys}^1,\,\vec{\ys}^2,\dots,\vec{\ys}^{\lambda}}]$.

For the long-term regime, which we 
define  by $k>\mbox{max}\{m_1, \dots,m_{\lambda-1}\}$, 
our analysis yields the following results:
\begin{itemize}
\item Decoherence functional in case  
$k>\mbox{max}\{m_1,m_2, \dots, m_{\lambda-1}\}$:
\begin{equation*}\hspace*{-9cm}
{\mathcal D}_{B,\,\rho_0}
[h_{\vec{\ys}^1,\,\vec{\ys}^2,\dots,\vec{\ys}^{\lambda}}\,,\,
h_{\vec{\zs}^1,\,\vec{\zs}^2,\dots,\vec{\zs}^{\lambda}}]
=
\end{equation*}
\begin{eqnarray}\label{result_[k>min_m_j]}
&=&
2^{-k-(m_1+\dots+m_{\lambda-1})}
\cdot\underbrace{\left(\,\prod_{j=1}^{k}
\prod_{i=1}^{\lambda}
\delta_{\ys^{j,i}}^{\zs^{j,i}}
\right)}_{{\rm diagonal}}\cdot
\underbrace{\left(\prod_{i=1}^{\lambda}
\delta{}_{\ys^{1,i}_{1:s_i-1}}^{\xs^i_{2:s_i}}
\right)}_{{\rm first \; shift}}\times
\nonumber\\
&&\quad\quad\quad\quad\quad\quad\quad \times
\underbrace{\left(\,\prod_{j=1}^{k-1}\prod_{i=1}^{\lambda}
\delta{}_{\ys^{j+1,i}_{1:s_i-1}}^{\ys^{j,i}_{2:s_i}}
\right)}_{{\rm step-by-step\ shift}}\cdot
\underbrace{\left(\prod_{i=1}^{\lambda}
\delta{}_{\ys^{k,i}_{1:s_i-k}}^{\xs^i_{k+1:s_i}}
\right)}_{k{\rm -th\ shift }}\\&&\quad\quad\quad\quad\quad\quad
\;\; +\,\, {\cal
  O}\Big(\frac{l+r-k}{2^{l-2(k^2+(1+m_1+m_2+\dots+m_{\lambda-1})k)}}\Big)\;
\nonumber
\end{eqnarray}
\item Entropy after $k$ iteration steps in case  
$k>\mbox{max}\{m_1,m_2, \dots, m_{\lambda-1}\}$:
\begin{equation}
H[\{ 
h_{\vec{\ys}^1,\,\vec{\ys}^2,\dots,\vec{\ys}^{\lambda}}\}] =
k+\sum_{i=1}^{\lambda-1}m_i\;+\;{\cal
  O}(\frac{(l+r-k)\log_2(l+r-k)}{2^{l-2(k^2+(1+m_1+m_2+\dots+m_{\lambda-1})k)}})\;.
\end{equation}
\end{itemize}
The interpretation of these results is similar to the special case $\lambda=2$
of the last section. Whereas in the short-term regime, $k<\mbox{min}\{m_1,
\dots, m_{\lambda-1}\}$, the entropy production rate is $\lambda$ bits per
iteration, in the long-term regime $k>\mbox{max}\{m_1,\dots,m_{\lambda-1}\}$,
the entropy production drops to $1$ bit per iteration, independently of the
values of the parameters $m_1,\dots,m_{\lambda-1}$, which determine the border
between the regimes. In the intermediate regime, the entropy production rate
decreases each time $k$ exceeds one of the values $m_1,\dots,m_{\lambda-1}$.

\section*{Acknowledgments}

We would like to thank Todd Brun for helpful
discussions. 
This work was supported in part by the European Union IST-FET project EDIQIP.


\begin{thebibliography}{10}


\bibitem{Gell-MannHartle1993}
M. Gell-Mann and J. B. Hartle, Phys.\ Rev.\ D {\bf 47}, 3345 (1993).

\bibitem{BrunHartle1999-PRD}
T.~A. Brun and J.~B. Hartle, Phys.\ Rev.\ D {\bf 60},  123503  (1999).

\bibitem{Griffiths1984}
R. Griffiths, J. Stat.\ Phys.\ {\bf 36},  219  (1984).

\bibitem{Omnes1988}
R. Omn\`{e}s, J. Stat.\ Phys.\ {\bf 53},  893, 933, 957  (1988).

\bibitem{Gell-MannHartle1990}
M. Gell-Mann and J.~B. Hartle,  in {\em Complexity, Entropy, and the Physics of
  Information}, edited by W.~H. Zurek (Addison Wesley, Redwood City, CA, 1990).

\bibitem{DowkerHalliwell1992}
H.~F. Dowker and J.~J. Halliwell, Phys.\ Rev.\ D {\bf 46},  1580  (1992).

\bibitem{Balazs1989}
N.~L. Balazs and A. Voros, Ann.\ Phys.\ {\bf 190},  1  (1989).

\bibitem{Saraceno1990}
M. Saraceno, Ann.\ Phys.\ {\bf 199},  37  (1990).

\bibitem{Arnold1968}
V.~I. Arnold and A. Avez, {\em Ergodic Problems of Classical Mechanics}
  (Benjamin, New York, 1968).

\bibitem{Berry1979}
M.~V. Berry, N.~L. Balazs, M. Tabor, and A. Voros, Ann.\ Phys.\ {\bf 122},  26
  (1979).

\bibitem{Weyl1950}
H. Weyl, {\em The Theory of Groups and Quantum Mechanics} (Dover, New York,
  1950).

\bibitem{Schack2000a}
R. Schack and C.~M. Caves, Applicable Algebra in Engineering, Communication and
  Computing (AAECC) {\bf 10},  305  (2000).

\bibitem{Alekseev1981}
V.~M. Alekseev and M.~V. Yakobson, Phys.\ Reports {\bf 75},  287
(1981).

\bibitem{Soklakov2000a}
A.~N. Soklakov and R. Schack, Phys.\ Rev.\ E {\bf 61},  5108  (2000).

\bibitem{Hartle1998}
J.~B. Hartle, Physica Scripta {\bf T76},  67  (1998).

\bibitem{BrunHartle1999-PRE}
T.~A. Brun and J.~B. Hartle, Phys.\ Rev.\ E {\bf 59},  6370  (1999).

\bibitem{Brun1995}
T.~A. Brun, Phys.\ Lett.\ A {\bf 206},  167  (1995).

\bibitem{Scherer2004b}
A.\ Scherer and A.~N.\ Soklakov, J.~Math.\ Phys.~46, 042108 (2005).

\bibitem{Soklakov2002}
A.~N.\ Soklakov and R.\ Schack, Phys.\ Rev.\ E {\bf 66}, 036212 (2002).

\end{thebibliography}
\end{document}